\documentclass[aps,prb,floatfix,amsmath,amssymb,preprint,eqsecnum,groupedaddress,showpacs,nofootinbib]{revtex4}

\usepackage[dvips]{graphicx}
\usepackage{dcolumn}
\usepackage{bm}

\usepackage{graphicx}
\usepackage{color}
\usepackage{feynmp}
\DeclareMathOperator{\im}{Im}

\newcommand{\D}{\mathcal{D}}
\newcommand{\Act}{\mathcal{S}}

\newcommand{\Ham}{\mathcal{H}}

\newcommand{\Mean}[1]{\left\langle#1\right\rangle}
\newcommand{\sub}[1]{_{\mathrm{#1}}}

\newcommand{\be}{\begin{equation}}
\newcommand{\ee}{\end{equation}}

\newcommand{\Order}[1]{\mathcal{O}\left(#1\right)}
\newcommand{\parder}[2]{\frac{\partial #1}{\partial #2}}
\newcommand{\parderat}[3]{{\left(\frac{\partial #1}{\partial #2}\right)}_{#3}}

\newcommand{\comm}[2]{[#1,#2]}
\newcommand{\bra}[1]{\langle #1|}
\newcommand{\ket}[1]{|#1\rangle}

\newcommand{\e}{\mathrm{e}}
\newcommand{\ii}{\mathrm{i}}
\newcommand{\dd}{\mathrm{d}}

\newcommand{\pcc}{+\mathrm{c.c.}}
\newcommand{\phc}{+\mathrm{h.c.}}
\newcommand{\punc}[1]{\;\mathrm{#1}}

\newcommand{\spaceitimeint}{\int \dd^d\xv\, \dd\tau\:}
\newcommand{\spacetimeint}{\int \dd^d\xv\, \dd t\:}

\newcommand{\Lv}{\mathbf{L}}
\newcommand{\kv}{\mathbf{k}}
\newcommand{\xv}{\mathbf{x}}
\newcommand{\nv}{\mathbf{n}}
\newcommand{\zerov}{\mathbf{0}}

\newcommand{\primafacie}{\textit{prima facie}}

\newcommand{\explicitcite}[1]{Ref.~\onlinecite{#1}}
\newcommand{\refeq}[1]{Eq.~(\ref{#1})}

\def\fmfL(#1,#2,#3)#4{\put(#1,#2){\makebox(0,0)[#3]{#4}}}

\begin{document}
\setlength{\unitlength}{1mm}

\title{Spin dynamics across the superfluid-insulator\\ transition of spinful bosons}

\author{Stephen Powell}
\affiliation{Department of Physics, Yale University, New Haven, CT
06520-8120}
\affiliation{Kavli Institute for Theoretical Physics,
University of California, Santa Barbara, CA 93106}

\author{Subir Sachdev}
\affiliation{Department of Physics, Harvard University, Cambridge MA
02138}

\begin{abstract}
Bosons with non-zero spin exhibit a rich variety of superfluid and
insulating phases. Most phases support coherent spin oscillations,
which have been the focus of numerous recent experiments. These spin
oscillations are Rabi oscillations between discrete levels deep in
the insulator, while deep in the superfluid they can be oscillations
in the orientation of a spinful condensate. We describe the
evolution of spin oscillations across the superfluid-insulator
quantum phase transition. For transitions with an order parameter
carrying spin, the damping of such oscillations is determined by the
scaling dimension of the composite spin operator. For transitions
with a spinless order parameter and gapped spin excitations, we
demonstrate that the damping is determined by an associated {\em
quantum impurity\/} problem of a localized spin excitation
interacting with the bulk critical modes. We present a
renormalization group analysis of the quantum impurity problem, and
discuss the relationship of our results to experiments on ultracold
atoms in optical lattices.
\end{abstract}





\date{February 2007\\[24pt]}

\maketitle

\section{Introduction}
\label{sec:intro}

An important frontier opened by the study of ultracold atoms has
been the investigation of Bose-Einstein condensates of atoms
carrying a nonzero total spin $F$. The condensate wavefunction then
has rich possibilities for interesting structure in spin space,
analogous to structure investigated earlier in superfluid $^3$He.
The dynamics of the atomic condensate is described by a
multi-component Gross-Pitaevski (GP) equation, and this allows for
interesting oscillations in the orientation of the condensate in
spin space. Such coherent spin oscillations have been observed in a
number of recent studies of $F=1$ and $F=2$ condensates
\cite{ketterle1,chapman1,kurn,chapman2,sengstock1,sengstock2}.

Coherent spin oscillations of a rather different nature are observed
in the presence of a strong optical lattice potential
\cite{bloch1,bloch2}. Here there is no Bose-Einstein condensate and
the ground state is a Mott insulator: each minimum of the optical
lattice potential traps a fixed number of atoms, and tunneling
between neighboring potential minima can be ignored. Now the spin
quantum number leads to a number of discrete atomic levels
associated with each minimum, with the energies determined by the
``cold collisional'' interactions between the atoms. The coherent
spin oscillations are then the Rabi oscillations between these
atomic levels \cite{bloch1,zhang}.

Note that the spin oscillations in the Mott insulator emerge from a
solution of the full quantum Schr\"odinger equation in the finite
Hilbert space of each potential minimum. In contrast, the
oscillations in the superfluid condensates
\cite{ketterle1,chapman1,kurn,chapman2,sengstock1,sengstock2} are
described by classical GP equations of motion obeyed by the
multicomponent order parameter, representing the collective
evolution of a macroscopic condensate of atoms.

It is the purpose of this paper to connect these distinct spin
oscillations to each other across the superfluid-insulator quantum
phase transition. The equilibrium properties of the
superfluid-insulator transition of spinful bosons are quite
complicated, and the very rich phase diagram has been explored
theoretically, both for \cite{demler1,kimura,fazio} $F=1$ and for
\cite{lewenstein} $F=2$. It is not our purpose here to shed further
light on the nature of this phase diagram, or on the possibilities
for the experimental realization of the various phases. Rather, we
will examine representative cases which display the distinct
possibilities in the evolution of coherent spin oscillations.

Deep in the superfluid, the spin oscillations are associated with
normal modes of the GP equations, about the superfluid state. Some
of these modes carry spin, and so will contribute to the spin
oscillations. If the superfluid state breaks spin rotation
invariance, then such a spin-carrying mode will be gapless.
Otherwise, the spin oscillations are gapped, ie, they occur at a
finite frequency. The dominant damping of these oscillations will
arise from the creation of low energy `phonon' excitations of the
superfluid. However, the gradient coupling to such `Goldstone' modes
will suppress the decay, and so one expects the oscillations to be
well-defined.

Deep in the Mott insulator, the discrete energy levels of each
potential minimum lead to undamped oscillations. Coupling between
these minima is expected to again lead to weak damping.

Assuming a second-order quantum phase transition between these
limits, we can expect that the spin oscillations will experience a
maximum in their damping at the critical point, and possibly even
cease to exist as well-defined modes. There are a plethora of low
energy excitations at the quantum critical point, and their coupling
to the spin modes is not constrained by the Goldstone theorem. The
primary purpose of this paper is to describe this enhanced damping
in the vicinity of the quantum critical point.

Our analysis shows that behavior of the spin oscillations falls into
two broad classes, and we will present a detailed analysis of a
representative example from both classes. The classes are: \\
({\bf A}) The order parameter for the quantum transition carries
non-zero spin. The spin excitation spectrum is gapless at the
quantum critical point, and the spin operator is characterized by
its scaling dimension. Typically, the spin operator is a composite
of the order parameter, and standard methods can be used to
determine its scaling dimension. The value of this scaling dimension
will determine the long-time decay of spin correlations, and hence
is a
measure of the damping. \\
({\bf B}) The order parameter of the superfluid-insulator transition
is spinless. In this case, it is likely that all excitations with
non-zero spin remain gapped across the quantum critical point. We
then have to consider the interaction of a single gapped spin
excitation with the gapless, spinless critical modes. Such a problem
was first considered in Refs.~\onlinecite{stv,japan} for the case of
gapped fermionic excitations. Here we will show that closely related
considerations apply also to the present case with gapped bosonic
excitations. The dispersion of the gapped excitation is argued to be
an irrelevant perturbation, and so to leading order one need only
consider the coupling of a localized spin excitation interacting
with the bulk critical modes. This gives the problem the character
of a {\em quantum impurity\/} problem. Depending upon whether the
localized-bulk coupling is relevant or irrelevant, we then have two
sub-cases. ({\bf B1}) If the coupling between the localized and bulk
excitations is relevant, then a renormalization group (RG) analysis
is necessary to understand the structure of spin correlations. A
brief account of this RG analysis was presented
earlier\cite{letter}, and here we will present further details
describing the new, non-trivial impurity fixed point controls the
long-time physics. ({\bf B2}) For the case of an irrelevant
localized-bulk coupling (which we will also present here), the
damping is controlled by scaling dimensions of the bulk theory, and
no new impurity dimensions are needed.
\begin{figure*}
\includegraphics{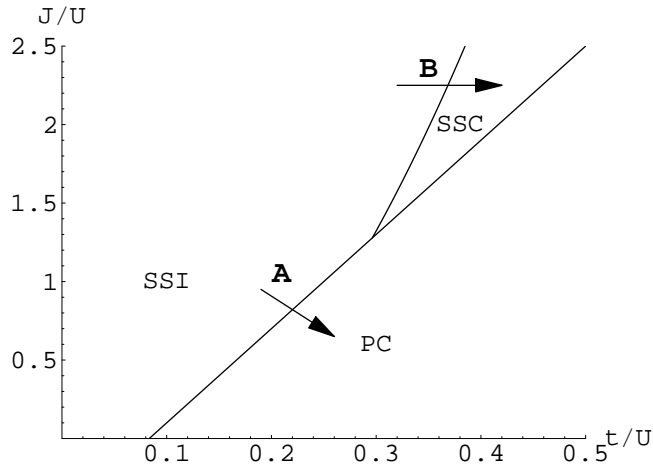}
\caption{\label{MeanFieldPhaseDiagram}Phase diagram for the
Hamiltonian in \refeq{Hamiltonian} calculated using the mean-field
theory of Section \ref{sec:model}. The three phases that are
included are the spin-singlet insulator (SSI), the spin-singlet
condensate (SSC), and the polar condensate (PC). The arrows labeled
`{\bf A}' and `{\bf B}' denote the two classes of transition defined
in Section \ref{sec:intro}. The calculation has been performed using
quantum rotors, corresponding to the canonical ensemble with the
filling factor an even integer. The horizontal and vertical axes
give the tunneling strength $t$ and the spin-dependent interaction
$J$, both in units of the spin-independent part of the interaction,
$U$.}
\end{figure*}

The representative examples we will consider in this paper will all
be drawn from the simplest case of $F=1$ bosons in an optical
lattice potential, with an {\em even\/} number, $N$, of bosons per
site \cite{demler1,kimura,fazio}. The mean-field phase diagram of a
model Hamiltonian for $F=1$ bosons is shown in
Fig.~\ref{MeanFieldPhaseDiagram}, along with the locations of the
quantum phase transitions of the two classes described above.
Further details on the phases and phase transitions appear in the
body of the paper. We expect that transitions with the more complex
order parameter categories possible for\cite{lewenstein} $F=2$ and
higher\cite{ryan}, will also fall into one of the categories we have
described above.

We will begin in Section~\ref{sec:model} by describing the model
Hamiltonian and a simple mean-field theory that can be used to treat
those phases with which we shall be concerned. We defer a more
thorough description of the symmetries and possible phases of the
model to Section~\ref{sec:fieldtheory}, where we consider the
continuum limit of the theory.

In Section~\ref{sec:PropertiesOfPhases}, we use this continuum
theory to describe the low-energy properties of the phases with
which we are concerned, focusing on the behavior of spin
excitations. We then turn, in Section~\ref{sec:qpt}, to the behavior
across transitions falling into the classes identified above. The
remaining sections treat these cases in more detail.

\section{Model and mean-field theory}
\label{sec:model}

\subsection{Model}
\label{sec:Hamiltonian}

Bosonic atoms trapped in an optical lattice potential, at
sufficiently low temperatures that all atoms occupy the lowest Bloch
band, are well described by the Bose-Hubbard model. (See, for
instance, \explicitcite{LewensteinReview} for a recent review.) The
extension of this model to the case where the bosons carry spin is
straightforward.  While most of our results apply for general spin
$F\neq 0$, we will treat explicitly the case $F=1$ and draw
attention to the generalizations when appropriate. The Hamiltonian
can then be written as \be \label{Hamiltonian} \Ham =
-t\sum_{\langle i,j\rangle} (b_{i\mu}^\dag b_{j\mu} + b_{j\mu}^\dag
b_{i\mu}) + \sum_i V(b_{i\mu}^\dag b_{i\mu}) + J\sum_i
|\Lv_i|^2\punc{,} \ee where summation over repeated spin indices
$\mu$ is implied throughout. The operator $b_{i\mu}$ annihilates a
boson at site $i$ with spin index  $\mu\in\{x,y,z\}$; this basis
will be the most convenient for our purposes.

The first term in $\Ham$ involves a sum over nearest-neighbor pairs
of sites, and allows the bosons to tunnel, with `hopping' strength
$t$. We will restrict attention to square and cubic lattices, but
the results can be straightforwardly generalized to other cases. The
function $V(n)$ contains an on-site spin-independent interaction and
the chemical potential, and can be written in the form \be V(n) =
\frac{1}{2}Un(n-1) - \mu n\punc{.} \ee

In the final term of \refeq{Hamiltonian}, $\Lv_i$ is the total
angular momentum on the site $i$, given by (for $F=1$) $L_{i,\rho} =
-\ii\epsilon_{\mu\nu\rho}b_{i\mu}^\dag b_{i\nu}$, where
$\epsilon_{\mu\nu\rho}$ is the completely antisymmetric tensor. For
$F=1$, this term is the most general quartic on-site spin-dependent
interaction. For spin $F$, the boson operator becomes a tensor of
rank $F$ and there are $F+1$ independent quartic interaction terms
corresponding to different contractions of the spin indices. We will
not include any direct interactions between spins on neighboring
sites, nor long-range polar forces between the atoms.

Suppose that $V(n)$ has its minimum near some even integer, $N$, and
the couplings are tuned so that the model is particle-hole symmetric
around this filling. Requiring this symmetry corresponds to
restricting consideration to the case of integer filling factor, and
is equivalent to using the canonical ensemble \cite{book}.

The spinless Bose-Hubbard model, with a single species of boson, has
a transition \cite{fwgf,book} from a Mott insulator when $U \gg t$
to a superfluid when $U \ll t$. When the bosons have spin, various
types of spin ordering are possible within both the insulator and
superfluid.

With an even number of particles per site, the simplest insulating
phase, the spin-singlet insulator (SSI), has a spin singlet on each
site. This will be favored energetically when $J>0$, and we will
concentrate on this case in the following. For $J<0$ or odd $N$, the
net spin on each site will be nonzero, and the system will be well
described by a quantum spin model, allowing various forms of spin
ordering within the lattice \cite{demler1}.

In a simple superfluid, the bosons condense, so that
$\Mean{b_{i\mu}}\neq 0$, breaking both gauge and spin symmetries.
For $J>0$, a so-called polar condensate (PC) is favored, with
$\Mean{b_{i\mu}} \propto \delta_{\mu z}$, where $z$ is an arbitrary
direction. As described in Ref.~\onlinecite{ryan}, a variety of
other condensates are, in general, also allowed.

For large enough $J$, a second variety of superfluid is possible
\cite{DemlerZhou}, within which single bosons have not condensed.
Instead, pairs of bosons condense, giving $\Mean{b_{i\mu}b_{i\mu}}
\neq 0$, which does not break spin-rotation symmetry. For this
state, the spin-singlet condensate (SSC), to be energetically
favorable, $J$ must be large enough to overcome the kinetic-energy
cost associated with pairing.

Referring back to the classes described in Section \ref{sec:intro}
and Fig.~\ref{MeanFieldPhaseDiagram}, we see that the transition
from SSI into PC is an example of class {\bf A}, in which the order
parameter carries nonzero spin, while the transition from SSI into
SSC is in class {\bf B}, with a spinless order parameter. We will
therefore mainly focus on these two phase transitions in the
following.

\subsection{Strong-pairing limit}
\label{SectionStrongPairingMF}

To provide a concrete, if qualitative, guide to the phase structure
of the particular model in \refeq{Hamiltonian}, we will implement a
mean-field theory capable of describing the phases of interest. This
will be similar to the approach of \explicitcite{fwgf} for the
spinless Bose-Hubbard model, where a mean-field is used to decouple
the hopping term.

Before describing this calculation, we will first use a very simple
perturbative calculation to give an approximate criterion for
condensation of boson pairs. In the limit of large $J/U$, an odd
number of bosons on any site is strongly disfavored, and we can deal
with a reduced Hilbert space of singlet pairs.

The effective tunneling rate $\tau$ for such pairs is given by $\tau
\sim t^2/v_1$, where $v_1 = U + 2J$ is the energy of the
intermediate state with a `broken' pair. The effective repulsion of
two pairs (ie, four bosons) on the same site is $\Upsilon \sim v_2 =
4U$.

We therefore arrive at the simple criterion $zt^2 \gtrsim U(U+2J)$
for the condensation of pairs, where $z$ is coordination number of
the lattice. This should be compared with the criterion $zt \gtrsim
U+2J$ for the condensation of single bosons \cite{fwgf,book}. These
two simple results will be confirmed, and the numerical prefactors
determined, by the mean-field analysis that follows (see Figure
\ref{MeanFieldPhaseDiagram}).

\subsection{Quantum rotor operators}
\label{SectionQuantumRotors}

\newcommand{\nh}{\hat{n}}

To simplify the mean-field calculation somewhat, we will use
$\mathrm{SO}(2)$ quantum rotor operators $\nh_i$ and $a_{i\mu}$ in
place of boson operators in the mean-field calculation. These
satisfy \be \comm{a_{i\mu}}{\nh_j} = \delta_{ij} a_{i\mu} \ee and
\be \comm{a_{i\mu}}{a_{j\nu}^\dag} = 0\punc{.} \ee This
simplification, which automatically incorporates particle-hole
symmetry, is convenient but inessential. The eigenvalues of $\nh_i$
are both positive and negative integers; they will be constrained to
physical values by the potential.

The full Hamiltonian, from \refeq{Hamiltonian}, can be written as
$\Ham = V - T$, where $V$ is the on-site interaction and $T$ is the
kinetic energy term. In the rotor formalism, $V$ is \be
\label{OnsiteV} V = \sum_i \left[U(\nh_i-N)^2 + J
|\Lv_i|^2\right]\punc{,} \ee where the term involving the chemical
potential has been absorbed by making the spin-independent
interaction explicitly symmetric about $N$. For simplicity, we will
take $N=2$ in the following. In the rotor formalism, the angular
momentum $\Lv_i$ is defined by its commutation relations with
$a_{i\mu}$, and the kinetic term is \be T = t\sum_{\langle
i,j\rangle} (a_{i\mu}^\dag a_{j\mu} \phc)\punc{.} \ee

First, consider the case when $t = 0$. Then the Hamiltonian is
simply a sum of terms acting on a single site, containing only the
commuting operators $\nh$ and $|\Lv|^2$. The ground state on each
site, which we label $\ket{2,L}$, is therefore an eigenstate both of
$\nh$, with eigenvalue $2$, and of $|\Lv|^2$, with eigenvalue
$L(L+1)$. For positive $J$, the ground state is a spin-singlet with
$L = 0$ and the lowest-lying `charged' excitations\footnote{Here,
and in the following, we use the term `charged' to refer to
excitations that change the particle number, assigning to particles
and holes charges of $+1$ and $-1$ respectively. Of course the
bosons have no electric charge and feel no long-range forces.} are
triplets with $L = 1$.

On the other hand, for negative $J$, the maximal value of $L$ is
favored, which will lead (once a small intersite coupling is
reinstated) to magnetic ordering. We therefore identify this latter
case with the NI phase; the level crossing that occurs when $J$
becomes negative will correspond in the thermodynamic limit to a
first-order transition out of the orderless SSI.

In the following we restrict to the case $J > 0$, to identify the
phase boundaries into the SSC and PC phases.

\subsection{Mean-field Hamiltonian}

We will proceed by choosing a mean-field (variational) ansatz that
incorporates the symmetry-breaking of the phases of interest. We
choose to do so by defining a mean-field Hamiltonian $\Ham\sub{mf}$,
whose ground state will be taken as the variational ansatz.

An appropriate mean-field Hamiltonian is \be \Ham\sub{mf} = V -
T_\psi - T_\Psi - T_\Phi\punc{,} \ee where $V$ is the same on-site
interaction as in \refeq{OnsiteV} and $T_\psi$ is the standard
mean-field decoupling of the hopping term, generalized to the case
with spin, \be T_\psi = \sum_i \left[ \psi_\mu a_{i\mu}^\dag +
\psi_\mu^* a_{i\mu}\right]\punc{,} \ee where $\psi_\mu$ is a
(c-number) constant vector, which will be used as a variational
parameter. The remaining terms allow for the possibility of a
spin-singlet condensate through the parameters $\Psi$ and $\Phi$:
\be T_\Psi = \sum_i\left[ \Psi a_{i\mu}^\dag a_{i\mu}^\dag + \Psi^*
a_{i\mu}a_{i\mu}\right]\punc{,} \ee and \be T_\Phi = \sum_{\langle
i,j\rangle}\left[ \Phi a_{i\mu}^\dag a_{j\mu}^\dag + \Phi^*
a_{i\mu}a_{j\mu}\right]\punc{,} \ee where the sum is over
nearest-neighbor pairs within the lattice.

\newcommand{\mf}{\mathrm{mf}}
\newcommand{\up}[1]{^{(#1)}}

We now use the ground state of $\Ham\sub{mf}$, which we denote
$\ket{\mf}$, as a variational ansatz and define \be
E\sub{mf}(\psi_\mu, \Psi, \Phi) = \bra{\mf}\Ham\ket{\mf}\punc{,} \ee
which should be minimized by varying the three parameters. If this
minimum occurs for vanishing values of all three parameters, then
$\ket\mf$ breaks no symmetries and the SSI phase is favored. A
nonzero value for $\psi_\mu$ at the minimum corresponds to PC, while
vanishing $\psi_\mu$ but nonzero values of $\Psi$ and/or $\Phi$
corresponds to SSC.

Since $\Ham\sub{mf}$ contains terms (within $T_\Phi$) that link
adjacent sites, it cannot be straightforwardly diagonalized, as in
the standard mean-field theory for the spinless Bose-Hubbard model.
To find the phase boundaries, however, we need only terms up to
quadratic order in the variational parameters, which can be found
using perturbation theory.

\subsection{Variational wavefunction}

To order zero in $\psi_\mu$, $\Psi$ and $\Phi$, we require the
ground state of $V$, \refeq{OnsiteV}. Assuming $U>0$ and $J>0$, this
is given by the simple product state \be \ket{\mf\up{0}} = \prod_j
\ket{2,0}_j\punc{.} \ee To first order, the ground state of
$\Ham\sub{mf}$ is \be \ket{\mf\up{1}} = \left(\frac{1}{v_1}T_\psi +
\frac{1}{v_2}T_\Psi +
\frac{1}{2v_1}T_\Phi\right)\ket{\mf\up{0}}\punc{,} \ee where $v_1 =
U + 2J$ and $v_2 = 4U$. (If rotors were not used in place of boson
operators, a similar but somewhat more complicated expression would
result.) All the physics incorporated in the mean-field ansatz is
visible at this order: The first term allows for a condensate of
single bosons, while the last two terms allow for a condensate of
spin singlets. The third term is necessary to allow these singlet
pairs to move around the lattice and make the SSC phase
energetically favorable.

Computing $E\sub{mf}$ to quadratic order in the variational
parameters (which requires the expression for the perturbed states
also to quadratic order) gives \be E\sub{mf} =
\begin{pmatrix}
\psi_\mu\\\Psi\\\Phi
\end{pmatrix}^\dagger
\begin{pmatrix}
\frac{2}{v_1}-\frac{4tz}{v_1^2}&0&0\\
0&\frac{2}{v_2}&-\frac{9tz}{v_1 v_2}\\
0&-\frac{9tz}{v_1 v_2}&\frac{3z}{2v_1}
\end{pmatrix}
\begin{pmatrix}
\psi_\mu\\\Psi\\\Phi
\end{pmatrix}\punc{.}
\ee The transition to PC occurs when the top-left element in the
matrix vanishes, while the transition to SSC occurs when the
determinant of the remaining block vanishes. This gives the criteria
$2 z t > v_1$ for PC and $27z t^2 > v_1 v_2$ for SSC, in agreement
with the simple considerations of Section
\ref{SectionStrongPairingMF}. The phase boundaries were shown in
Figure \ref{MeanFieldPhaseDiagram}.

Note that it is in fact necessary to continue the expansion to
fourth order to determine the direction of the vector $\psi_\mu$
when it is nonzero. This calculation has been performed in
\explicitcite{Tsuchiya}, where the possibility of SSC was not
incorporated and the PC phase was found to be favored, as expected.

In principle it is also necessary to continue the expansion to
higher order to investigate the competition between PC and SSC in
the region where both are possible. Simple energetic considerations
suggest, however, that condensation of single bosons in the PC phase
will dominate over condensation of pairs, and this has been assumed
in Figure \ref{MeanFieldPhaseDiagram}.

\section{Symmetries and phases}
\label{sec:fieldtheory}

\subsection{Continuum action}
\label{sec:ContinuumAction}

To describe the low-energy excitations of this model, we will derive
a continuum field theory that captures the physics near to zero
momentum. This assumes the absence of antiferromagnetic ordering of
the spins, for example; it is chosen to be appropriate to the phases
with which we are concerned.

The action that results is in fact completely determined by the
symmetries of the model, but a formal derivation is possible by
analogy to the standard (spinless) Bose-Hubbard model
\cite{fwgf,book}. One first writes the partition function as a path
integral and then decouples the hopping term using a site-dependent
field $\psi_\mu$. Perturbation theory in $\psi_\mu$ can then be used
to eliminate all excitations above the ground state.

The final form of the action has the same $\mathrm{U}(1)$ phase and
$\mathrm{SU}(2)$ spin symmetry as the original Hamiltonian. Since
the parameters have been chosen to give particle-hole symmetry, it
can be written in a relativistic form: \be \label{Actpsi} \Act_\psi
= \spacetimeint \left( -\bar{\psi}_\mu \partial^2 \psi_\mu +
r\,\bar{\psi}_\mu \psi_\mu \right) + \Act_\psi^{(4)}
+\,\cdots\punc{.} \ee Note that the action is completely
relativistic and the derivative $\partial$ acts in $D=d+1$
dimensions: $\partial^2 = \nabla^2 - \partial_t^2$. The quartic
interaction contains two terms: \be \label{Actpsi4} \Act_\psi^{(4)}
= \spacetimeint \left(\frac{u}{4}\,\bar{\psi}_\mu \psi_\mu
\bar{\psi}_\nu \psi_\nu + \frac{v}{4}\,\bar{\psi}_\mu \bar{\psi}_\mu
\psi_\nu \psi_\nu\right)\punc{.} \ee The first term has
$\mathrm{O}(6)$ symmetry, while the second term, which vanishes if
$J = 0$ in $\Ham$, breaks this down to $\mathrm{SU}(2)$. We will be
interested in the case $v<0$, for that yields a superfluid PC state.

For higher spin $F$, the action has a similar form, with the field
$\psi$ becoming a tensor of rank $F$. The quadratic part of the
action is unchanged, but the quartic term now involves the $F+1$
distinct scalar contractions of the field.

\subsection{Symmetries}
\label{sec:Symmetries}

The action $\Act_\psi$ has full spatial symmetry, as well as the
following (global) internal symmetries.

\begin{itemize}

\item Spin rotation ($S$), under which $\psi_\mu$ is a vector

\item Phase rotation ($\Phi$): $\psi_\mu \rightarrow \psi_\mu \e^{\ii\phi}$

\item Spin or phase inversion ($I$): $\psi_\mu \rightarrow -\psi_\mu$

\item Time reversal ($T$): $\bar\psi_\mu \leftrightarrow \psi_\mu$, $\partial_\tau \rightarrow -\partial_\tau$

\end{itemize}

These symmetries are not independent, and ground states breaking
some of these symmetries necessarily break others. Conversely,
unbroken spin-rotation symmetry, for instance, implies unbroken
spin-inversion symmetry, which we denote \be S \implies I\punc{.}
\ee Similarly, we have \be S \implies T \ee and \be \Phi \implies
I\punc{.} \ee Note that $\Phi$ does not imply $T$, so it is possible
to have broken time-reversal symmetry while maintaining
phase-rotation symmetry. The identification of spin and phase
inversion as the single operation $I$ implies that it is impossible
to break one without breaking the other \cite{DemlerZhou}.

\subsection{Observables}
\label{sec:Observables}

To connect the predictions of the continuum field theory to physical
observables, we must relate these to the field $\psi_\mu$.

Since angular momentum $\Lv$ is a pseudovector, symmetry requires
\be \label{L:field} L_\rho \sim
\ii\epsilon_{\mu\nu\rho}\bar{\psi}_\mu \psi_\nu\punc{.} \ee If
time-reversal symmetry $T$ is unbroken, we therefore have
$\Mean{\Lv} = 0$. (Note that unbroken $I$ is compatible with a
nonzero $\Mean{\Lv}$, since $\Lv$ is a pseudovector.)

In experiments with ultracold atoms, the population of each
individual hyperfine state is also potentially measurable. We could
consider, for instance, the operator $b_{iz}^\dag b_{iz}$, which
counts the number of bosons in the $\mu = z$ spin state (at site
$i$). It is more convenient to define instead \be Q_{zz} =
b_{iz}^\dag b_{iz} - \frac{1}{3}b_{i\rho}^\dag b_{i\rho}\punc{,} \ee
which measures the `population imbalance' towards this state and has
zero expectation value in a state without spin ordering.

As suggested by the notation, $Q_{zz}$ is in fact a component of a
(symmetric, traceless) second-rank tensor \be \label{DefineQ}
Q_{\mu\nu} = \frac{1}{2}\left(b_{i\mu}^\dag b_{i\nu}+b_{i\nu}^\dag
b_{i\mu}\right) - \frac{1}{3}\delta_{\mu\nu}b_{i\rho}^\dag
b_{i\rho}\punc{.} \ee In terms of the continuum fields, symmetry
implies \be Q_{\mu\nu} \sim \frac{1}{2}\left(\bar\psi_\mu
\psi_\nu+\bar\psi_\nu \psi_\mu\right) -
\frac{1}{3}\delta_{\mu\nu}\bar\psi_\rho\psi_\rho\punc{.} \ee

\subsection{Classification of phases}
\label{sec:Phases}

We now list the phases described by the continuum theory of Section
\ref{sec:ContinuumAction}, which allows for the breaking of spin and
phase symmetries, but preserves the full spatial symmetry of the
original lattice. The relevant connected correlation functions are
the following: \be
\begin{aligned}
\Mean{\psi_\mu} &= \varphi_0 n_\mu + \varphi\sub{f} \nu_\mu\\
\Mean{\bar{\psi}_\mu \psi_\nu}_c &= \psi_0^2 \frac{\delta_{\mu\nu}}{3} + \psi_1^2 \alpha_{\mu\nu} + \psi\sub{f}^2 \ii\epsilon_{\mu\nu\rho}n_\rho\\
\Mean{\psi_\mu \psi_\nu}_c &= \Psi_0 \frac{\delta_{\mu\nu}}{3} +
\Psi_1 \alpha_{\mu\nu} + \Psi\sub{f} \nu_\mu \nu_\nu
\end{aligned}
\label{MFAnsatz} \ee where $\varphi_{0,\mathrm{f}}$,
$\psi_{0,1,\mathrm{f}}$, and $\Psi_{0,1,\mathrm{f}}$ are scalar
parameters. The unit vector $n_\mu$ is arbitrary and
$\alpha_{\mu\nu}$ and $\nu_\mu$ are defined as \be \alpha_{\mu\nu} =
n_\mu n_\nu - \frac{1}{3}\delta_{\mu\nu} \ee and \be \nu_\mu =
\frac{1}{\sqrt{2}}(n_{1\mu} + \ii n_{2\mu})\punc{,} \ee where
$\nv_1$ and $\nv_2$ are mutually orthogonal unit vectors satisfying
$\nv_1\times \nv_2 = \nv$.

\begin{table*}
\begin{tabular}{|l|c|c|}
\hline
State & Broken symmetries & Nonzero parameters\\
\hline \hline
Spin-singlet insulator (SSI) & None & $\psi_0$\\
Spin-singlet condensate (SSC) & $\Phi$ & $\psi_0$, $\Psi_0$\\
\hline
Nematic insulator (NI) & $S$ & $\psi_0$, $\psi_1$\\
Strong-coupling pairing (SCP) & $\Phi$, $S$ & $\psi_0$, $\Psi_1$, $\Psi_0$, $\psi_1$\\
Polar condensate (PC) & $I$, $\Phi$, $S$ & $\psi_0$, $\varphi_0$, $\Psi_1$, $\Psi_0$, $\psi_1$\\
\hline
Ferromagnetic insulator (FI) & $S$, $T$ & NI + $\psi\sub{f}$\\
Ferromagnetic SCP (FSCP) & $\Phi$, $S$, $T$ & SCP + $\psi\sub{f}$, $\Psi\sub{f}$ \\
Ferromagnetic condensate (FC) & $I$, $\Phi$, $S$, $T$ & PC + $\varphi\sub{f}$, $\psi\sub{f}$, $\Psi\sub{f}$\\
\hline
\end{tabular}
\caption{The possible states described by the continuum theory of
Section \ref{sec:ContinuumAction}. The notation for the symmetries
is described in Section \ref{sec:Symmetries} and the nomenclature
for the states follows that used by \explicitcite{DemlerZhou}. Note
that the parameter $\psi_0$ breaks no symmetries and is therefore
nonzero in every state.} \label{MFStatesTable}
\end{table*}

All of the states allowed by this symmetry analysis are summarized
in Table \ref{MFStatesTable}. Note that each state with $S$ broken
has a corresponding state with $T$ also broken; these are states
with at least one of $\varphi\sub{f}$, $\psi\sub{f}$ and
$\Psi\sub{f}$ nonzero. Using \refeq{L:field}, we see that they are
ferromagnetically ordered and have $\Mean{\Lv}$ parallel to $\nv$.

Table \ref{MFStatesTable} is based on simple symmetry considerations
and not every state listed will be possible in any particular
physical realization. For example, the SCP phase appears, on the
basis of energetic considerations, to be inevitably unfavorable
compared to PC. Conversely, there are other possibilities for
ordering that are not incorporated in Table \ref{MFStatesTable},
such as breaking of spatial (lattice) symmetries \footnote{This is
particularly important in Mott-insulating states of fermions, but
may be less prominent in bosonic systems, where Bose enhancement
encourages neighboring sites to have ferromagnetically aligned
spins.}.

Since it is our purpose here to describe examples of the various
classes identified in Section \ref{sec:intro}, we will restrict our
attention to a handful of phases and the transitions between them.
Specifically, we will be interested in the three phases identified
in Section \ref{sec:model}: SSI, SSC, and PC, allowing for the
possibility of both spin and phase ordering.

\section{Properties of phases}
\label{sec:PropertiesOfPhases}

The various phases in Figure \ref{MeanFieldPhaseDiagram} have
different symmetries and low-energy excitations. We will now outline
the properties of these phases, focusing in each case on the
response to probes coupling to the angular momentum, $\Lv$, and to
$Q_{\mu\nu}$, defined in \refeq{DefineQ}. These are described by the
correlation functions \be \Pi_\Lv(\xv,t) \sim \Mean{\mathcal{T}_t \:
\Lv(\xv,t) \cdot \Lv(\zerov,0)} \ee and \be \Pi_Q(\xv,t) \sim
\Mean{\mathcal{T}_t \:
Q_{\mu\nu}(\xv,t)Q_{\mu\nu}(\zerov,0)}\punc{,} \ee where
$\mathcal{T}_t$ denotes time-ordering.

The explicit calculations will be carried out two spatial
dimensions, but most of the qualitative conclusions will also apply
in three dimensions.

\subsection{Spin-singlet insulator}

The SSI phase is a `featureless' insulator without spin or phase
ordering. All quasiparticle excitations are gapped, ie, they occur
at finite energy above the ground state.

The phase can be further divided according to the lowest-energy
`charged' excitation. Throughout most of the phase, individual
particle and hole excitations, described by the field $\psi_\mu$,
will have the smallest gap, but in a small region relatively close
to the transition to SSC, bound singlet pairs will move to lower
energy. (It is these excitations that condense across the transition
to SSC, as described below, in Section \ref{SSC}.) We will describe
the former case here and return to the latter in Section
\ref{SSInearSSC}.

In the absence of any condensate, the appropriate action is simply
that given in \refeq{Actpsi}, which we write, with $r = \lambda^2$,
as \be \label{SSIaction} \Act\sub{SSI} =
\spacetimeint\left[\bar{\psi}_\mu (-\partial^2+\lambda^2) \psi_\mu +
\frac{u}{4}\,\bar{\psi}_\mu \psi_\mu \bar{\psi}_\nu \psi_\nu +
\frac{v}{4}\,\bar{\psi}_\mu \bar{\psi}_\mu \psi_\nu \psi_\nu
+\,\cdots\right]\punc{.} \ee Particles and holes are described by
the same field $\psi_\mu$, with gap $\lambda$ and two distinct
quartic interactions, with coefficients $u$ and $v$.

Perturbation theory in $u$ and $v$, which we take to be on the same
order, will be used to describe the low-energy properties of this
phase. To do so, we first define the free propagator for the field
$\psi_\mu$: \be G^\psi_0(\omega, \kv) = \frac{1}{-\omega^2 + k^2 +
\lambda^2}\punc{.} \ee Because of relativistic invariance, the
results are given below for the case $\kv = \zerov$; the
corresponding expressions for nonzero momentum are given by the
replacement $\omega^2 \rightarrow \omega^2 - k^2$.

\subsubsection{Self energy}
\label{SSISelfEnergy}

First, we describe the self energy, for which the lowest-order
`tadpole' diagram is \be \label{SSItadpoleDiagram} \Sigma^\psi_1 =
\;
\parbox{30mm}{
\begin{picture}(30,20)
\put(0,0){\includegraphics{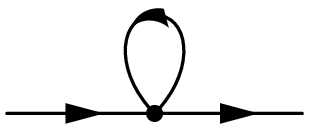}}

\end{picture}
} \ee (where the vertex represents some linear combination of $u$
and $v$). This diagram does not depend on the frequency or momentum
carried by the external line and so simply contributes a constant
that renormalizes the gap $\lambda$.

The lowest-order diagram that depends on the external momentum is
\be \label{SSIfirstDiagram} \Sigma^\psi_2 = \;
\parbox{30mm}{
\begin{picture}(30,20)
\put(0,0){\includegraphics{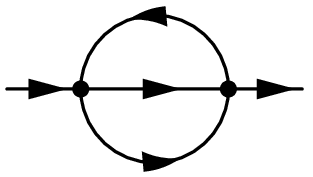}}

\end{picture}}\punc{.}
\ee The physical interpretation of this diagram is as the decay of a
particle, given sufficient energy, into a hole (described the top
line, with the reversed propagation direction) and two particles. It
is therefore clear that this diagram will make no contribution to
the decay rate for a particle unless its energy exceeds $3\lambda$.

\begin{figure*}
\includegraphics{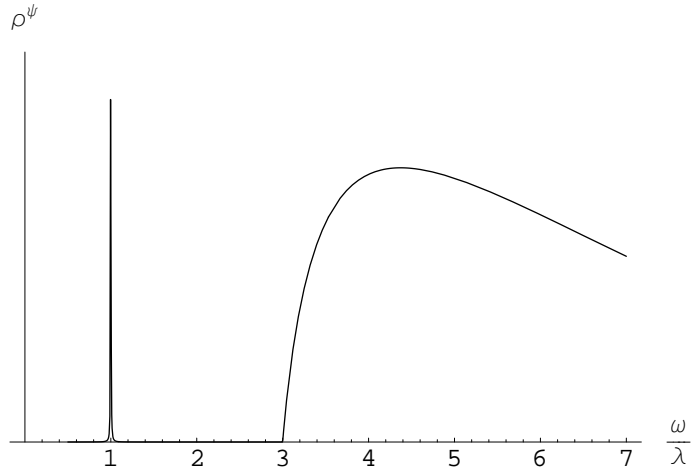}
\caption{\label{fig:SSIrho}The spectral weight $\rho^\psi$ in SSI,
calculated up to second order in the couplings $u$ and $v$ in $d=2$.
The delta-function peak at $\omega = \lambda$ (which has
artificially been given a small but nonzero width) describes the
stable particle excitation of the field $\psi_\mu$. For $\omega >
3\lambda$, there is continuum of three-particle excitations.}
\end{figure*}

This interpretation is clarified by using the spectral
representation; the `spectral weight' is given by \be
\label{SpectralWeightDefinition} \rho^\psi(\omega) = 2\lim_{\eta
\rightarrow 0} \im G^\psi(\omega + \ii \eta)\punc{,} \ee where
$G^\psi$ is the full propagator. This can be calculated numerically,
and is shown in Figure \ref{fig:SSIrho}. There is a delta-function
peak\footnote{We note here a subtlety regarding this plot and those
in the remainder of Section \ref{sec:PropertiesOfPhases}. The full
series of diagrams, including those shown in
\refeq{SSItadpoleDiagram} and \refeq{SSIfirstDiagram}, cause a
renormalization of the gap $\lambda$ (and hence movement of the
features in $\rho^\psi$) away from its bare value, ie, the value
appearing explicitly in the action. In this plot (and the ones that
follow), $\lambda$ should be interpreted as meaning the renormalized
value, rather than the bare value, and it is for this reason that
the peak appears precisely at $\omega = \lambda$.} at $\omega =
\lambda$ (a small width has been manually added to make it visible
on the plot), and a three-body continuum, resulting from
$\Sigma^\psi_2$, appears at $\omega = 3\lambda$. The spectral weight
above this 3-body threshhold can be estimated from a
non-relativistic theory of excitations above the gap, and yields
$\rho^\psi (\omega) \sim (\omega-3\lambda)^{d-1}$ for $\omega$ just
above $3 \lambda$.

\subsubsection{Spin response}
\label{sec:SSIspinresponse}

To describe the spin-response functions, we define \be
\label{PiFromPsi} \Pi_{\mu\nu,\rho\sigma} \sim \Mean{\mathcal{T}_t\:
\bar{\psi}_\mu(\xv,t) \psi_\nu(\xv,t) \bar{\psi}_\rho(\zerov,0)
\psi_\sigma(\zerov,0)}\punc{,} \ee in terms of which, \be \Pi_\Lv =
\Pi_{\mu\nu,\mu\nu} - \Pi_{\mu\nu,\nu\mu} \ee and \be \Pi_Q =
\frac{1}{2}\left(\Pi_{\mu\nu,\mu\nu} + \Pi_{\mu\nu,\nu\mu}\right) -
\frac{1}{3} \Pi_{\mu\mu,\nu\nu}\punc{.} \ee

\begin{figure*}
\includegraphics{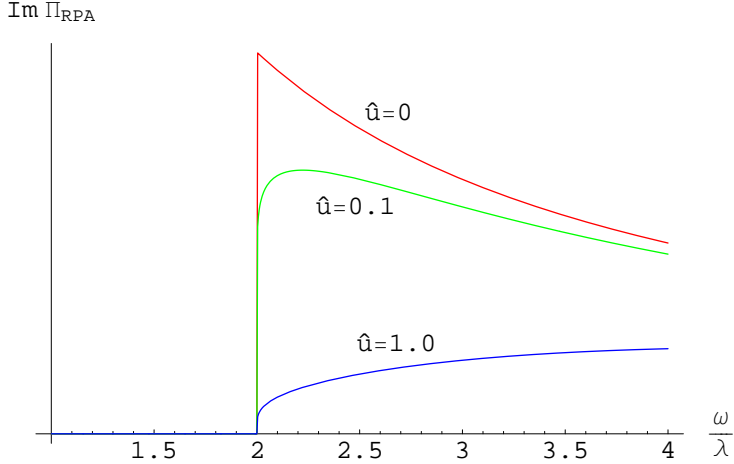}
\caption{\label{fig:SSIpi}The imaginary part of the spin-response
function $\Pi\sub{RPA}$ in SSI in $d=2$, calculated using the
random-phase approximation. The red line, with zero coupling, gives
the one-loop diagram $\Pi_0$, while the other two lines give the
results for nonzero coupling $\hat u$. This coupling constant is a
linear combination of $u$ and $v$, the two couplings appearing in
the action $\Act\sub{SSI}$, \refeq{SSIaction}.}
\end{figure*}

The lowest-order diagram contributing to $\Pi_{\mu\nu,\rho\sigma}$
is given by the `polarization bubble' \be \Pi_0 = \;
\parbox{30mm}{
\begin{picture}(30,20)
\put(0,0){\includegraphics{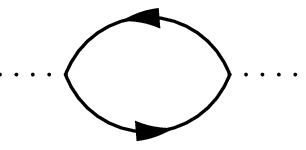}}

\end{picture}
}\punc{,} \ee which contains no interaction vertices. The dotted
lines represent insertions of $\bar{\psi}_\mu\psi_\nu$; taking into
account only this diagram gives spin-dependence
$\Pi_{\mu\nu,\rho\sigma}\propto \delta_{\mu\sigma}\delta_{\nu\rho}$.
Performing the integration over the loop momentum in $d=2$ gives the
simple result \be \Pi_0 = \frac{1}{8\pi \omega} \log
\frac{2\lambda+\omega}{2\lambda-\omega}\punc{,} \label{ss1} \ee
where $\omega$ is the frequency carried by the external lines. The
imaginary part of this function is shown as the red line in Figure
\ref{fig:SSIpi}, and has a discontinuity at the threshold, as
expected from Eq.~(\ref{ss1}). In general $d$, the behavior just
above threshold for the imaginary part is $(\omega-2
\lambda)^{(d-2)/2}$.

Apart from insertions of the simple tadpole diagram $\Sigma^\psi_1$
into one of the propagators, which can be accounted for by a
renormalization of the gap $\lambda$, the only diagrams of first
order in the couplings $u$ and $v$ have the form \be \Pi_1 = \;
\parbox{30mm}{
\begin{picture}(30,20)
\put(0,0){\includegraphics{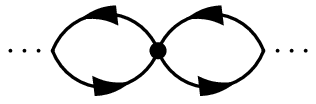}}

\end{picture}
}\punc{.} \ee Again, the vertex represents some linear combination
of $u$ and $v$, which we write $\hat u$, and which depends on the
particular response function of interest. Simple algebra in the spin
indices gives $\hat u = \frac{u}{2}-v$ for $\Pi_\Lv$ and $\hat u =
\frac{u}{2} + v$ for $\Pi_Q$. In SSI, these two response functions
therefore have similar behavior, but are described by different
combinations of the coupling constants.

The diagram $\Pi_1$ is simply given by $-\hat{u}\Pi_0^2$ and in fact
forms the second term of a geometric series. This set of diagrams
can be summed, leading to the `random-phase approximation' for the
response function, \be \Pi\sub{RPA} = \frac{\Pi_0}{1 + \hat{u}
\Pi_0}\punc{.} \ee In Figure \ref{fig:SSIpi}, the imaginary part of
this function is plotted in $d=2$, for three different values of the
coupling $\hat u$. It is nonzero only for $\omega > 2\lambda$, the
energy required to create a particle-hole pair. The discontinuity
found earlier at threshold in Eq.~(\ref{ss1}) is now suppressed
logarithmically by the RPA corrections: and for small $x = \omega -
2\lambda>0$, the singularity is of the form $1/\log x$.

\subsection{SSI near SSC}
\label{SSInearSSC}

As the spin-dependent interaction increases and the SSC phase is
approached, a bound state composed of a singlet pair of bosons
becomes energetically favorable. At the transition to SSC, the
singlet pairs condense into a superfluid with no spin ordering.

To describe the approach to this transition, we start with the
action $\Act\sub{SSI}$ and introduce the field $\Psi \sim \psi_\mu
\psi_\mu$, by a Hubbard-Stratonovich decoupling of the quartic
interaction $v$. The field $\Psi$ describes the singlet pairs and
will condense across the transition. It is described by the action
\be \label{ActPsiSSI} \Act_\Psi = \spacetimeint\left(|\partial
\Psi|^2 + r_\Psi |\Psi|^2 + \frac{u_\Psi}{4}|\Psi|^4 +
\,\cdots\right)\punc{.} \ee The full action has the form \be
\label{SSInearSSCaction} \Act\sub{SSI}' = \Act_\psi + \Act_\Psi +
\frac{g_\psi}{2}\spacetimeint\left( \bar\Psi \psi_\mu \psi_\mu +
\Psi \bar\psi_\mu \bar\psi_\mu \right)\punc{.} \ee (The introduction
of the field $\Psi$ renormalizes the coupling constants within
$\Act_\psi$. Here and throughout, we will simplify the notation by
retaining the same symbols for these renormalized quantities.)

\subsubsection{Self energy}

The diagrams shown in Section \ref{SSISelfEnergy}, coming from
$\Act_\psi$, will still contribute to the self energy near to SSC.
As seen above, however, these diagrams are important only for
$\omega > 3 \lambda$, whereas new diagrams coming from coupling to
the pair field $\Psi$ will contribute at lower frequencies.

Using a double line for the propagator of the $\Psi$ field, the
first diagram is \be \label{ParticleHolePairBubble} \Sigma_1^\psi =
\;
\parbox{30mm}{
\begin{picture}(30,20)
\put(0,0){\includegraphics{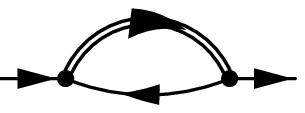}}

\end{picture}}\punc{,}
\ee in which a particle decays into a hole plus a pair. The vertices
correspond to factors of $g_\psi$. The threshold for this process is
clearly $\lambda + \lambda_\Psi$, where $\lambda$ is the gap to the
particle and hole excitation as before, and $\lambda_\Psi =
\sqrt{r_\Psi}$ is the gap to the pair excitation. This excitation
therefore becomes important at low energies when $\lambda_\Psi <
2\lambda$, which is simply the condition that a bound state exists
below the two-particle continuum.

\begin{figure*}
\includegraphics{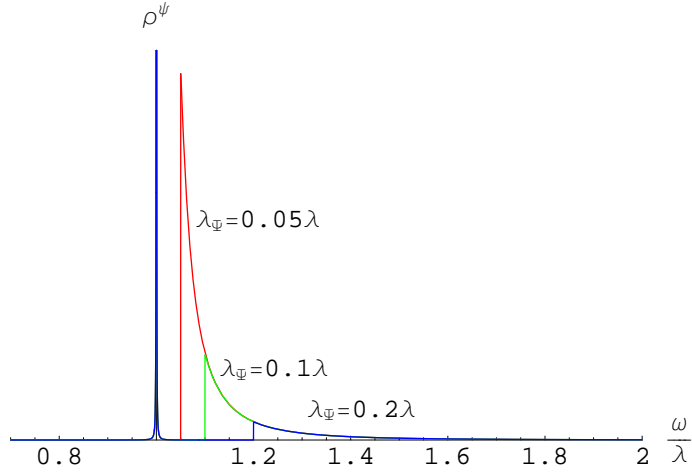}
\caption{\label{fig:SSISSCrho}The spectral weight $\rho^\psi$ in SSI
near the transition to SSC, calculated up to order $g_\psi^2$,
plotted for three different values of $\lambda_\Psi$, the gap to
pair excitations. The coupling strength is $g_\psi = 0.1$. The peak
at $\omega = \lambda$ (which is present for all values of
$\lambda_\Psi$ and has artificially been given a small but nonzero
width) describes the stable particle and hole excitation of the
field $\psi_\mu$. For $\omega > \lambda + \lambda_\Psi$, there is
continuum of excitations, corresponding physically to the conversion
of a particle to a pair plus a hole.}
\end{figure*}

The diagram can be evaluated in $d=2$ to give \be \Sigma_1^\psi =
\frac{g_\psi^2}{8\pi \omega}\log \frac{\lambda + \lambda_\Psi +
\omega}{\lambda + \lambda_\Psi - \omega}\punc{,} \ee and the
corresponding spectral weight is shown in Figure
\ref{fig:SSISSCrho}; the structure of this threshold singularity is
similar to Eq.~(\ref{ss1}), and as was the case there, in general
$d$ we have a singularity $\sim
(\omega-\lambda-\lambda_\Psi)^{(d-2)/2}$. A continuum of excitations
appears for $\omega
> \lambda + \lambda_\Psi$, as expected. As the transition to SSC is
approached, $\lambda_\Psi$ becomes smaller and the edge of the
continuum approaches the peak at $\omega = \lambda$. The
perturbation expansion used here breaks down as
$\lambda_\Psi\rightarrow 0$ and a more sophisticated RG calculation,
described in Section \ref{SectionRGforSSC}, is required.

\subsubsection{Spin response}

Since $\Psi$ is a spin singlet, it gives no direct contribution to
the spin response, which is therefore given by \refeq{PiFromPsi}, as
before. The presence of the bound state, however, allows for new
diagrams that contribute to the response function
$\Pi_{\mu\nu,\rho\sigma}$.

One such diagram is \be \label{Pi1a} \Pi^{(1a)} =\;
\parbox{30mm}{
\begin{picture}(30,20)
\put(0,0){\includegraphics{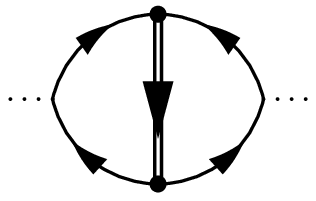}}

\end{picture}
}\punc{,} \ee which is the first term in the series \be
\parbox{20mm}{
\begin{picture}(20,15)
\put(0,0){\includegraphics{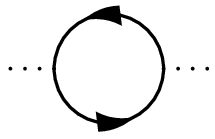}}

\end{picture}
} +
\parbox{25mm}{
\begin{picture}(25,15)
\put(0,0){\includegraphics{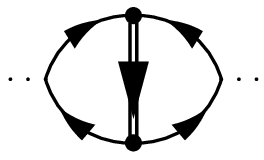}}

\end{picture}
} +
\parbox{30mm}{
\begin{picture}(30,15)
\put(0,0){\includegraphics{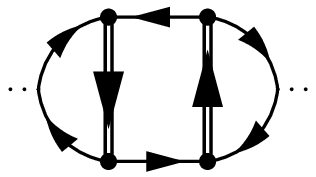}}

\end{picture}
} +
\parbox{40mm}{
\begin{picture}(40,15)
\put(0,0){\includegraphics{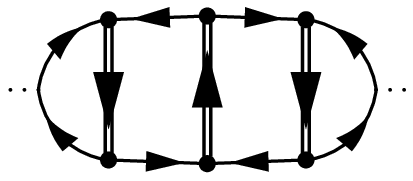}}

\end{picture}
} + \cdots \ee For $x = \omega - 2\lambda$ positive but very small,
the pair propagator can be replaced by a constant. Summing over
these diagrams will therefore lead to a similar $1/\log x$
singularity to that described above in Section
\ref{sec:SSIspinresponse}. This replacement is only valid for $x <
\lambda_\Psi$. For $\omega > 2\lambda + \lambda_\Psi$, a 3-particle
threshold singularity is found, similar to that discussed below
Eq.~(\ref{SpectralWeightDefinition}). Qualitatively different
behavior will result as the transition to SSC is approached and
$\lambda_\Psi \rightarrow 0$. This will be addressed below in
Section \ref{SectionRGforSSC}.

\subsection{Spin-singlet condensate}
\label{SSC}

In the SSC phase, singlet pairs of bosons form a condensate, giving
a superfluid with no spin ordering. This will only occur when the
spin-dependent interaction is strong enough to overcome the extra
kinetic-energy cost of having bound singlet pairs. This requires
sufficiently large $J/U$, as shown in Figure
\ref{MeanFieldPhaseDiagram}.

The perturbation expansion used in Section \ref{SSInearSSC} is not
applicable here, and we must instead expand about the new ground
state, with a condensed $\Psi$ field. (A similar approach can be
used to describe the condensed phase of the spinless Bose gas
\cite{Popov}.)

We first write $\Psi$ in terms of amplitude and phase as \be
\label{SSCphaserep} \Psi = \Psi_0 \e^{\ii \theta'}\punc{,} \ee where
$\Psi_0$ and $\theta'$ are both real. For simplicity, we treat the
amplitude of the $\Psi$ field as a constant, ignoring the gapped
amplitude modes. (This is appropriate sufficiently far from the
transition to SSI, where the gap is large.) With this
parametrization, $\Act_\Psi$, given by \refeq{ActPsiSSI}, can be
rewritten as the action of a free, gapless field: \be
\label{SSCactionPsi} \Act_\Psi = \spacetimeint \frac{1}{2}(\partial
\theta)^2\punc{,} \ee with the definition $\theta =
\sqrt{2}\Psi_0\theta'$. Physically, $\theta$ is interpreted as the
Goldstone mode corresponding to the broken phase symmetry in SSC.

In dealing with $\psi_\mu$, it is convenient to take out a factor of
the condensate phase by writing $\psi_\mu = \varphi_\mu \e^{\ii
\theta'/2}$. Then, since the condensate has broken phase-rotation
invariance, we rewrite the field $\varphi_\mu$ in terms of real and
imaginary parts, \be \varphi_\mu = \frac{1}{\sqrt{2}} (\varphi_\mu^R
+ \ii\varphi_\mu^I)\punc{.} \ee

In terms of the new fields $\theta$, $\varphi_\mu^R$ and
$\varphi_\mu^I$, the action becomes\footnote{It is straightforward
to show that the Jacobian associated with the change of variables is
equal to a constant.}
\begin{multline}
\label{SSCaction} \Act\sub{SSC} = \spacetimeint
\bigg\{\frac{1}{2}(\partial \theta)^2 + \frac{1}{2}\varphi_\mu^R
(-\partial^2 + r + g^\psi\Psi_0/2) \varphi_\mu^R
+ \frac{1}{2}\varphi_\mu^I (-\partial^2 + r - g^\psi\Psi_0/2) \varphi_\mu^I\\
+ \frac{u}{16}\,(\varphi_\mu^R \varphi_\mu^R + \varphi_\mu^I
\varphi_\mu^I)(\varphi_\nu^R \varphi_\nu^R + \varphi_\nu^I
\varphi_\nu^I) + \frac{v}{16}\,\left[(\varphi_\mu^R \varphi_\mu^R -
\varphi_\mu^I \varphi_\mu^I)(\varphi_\nu^R \varphi_\nu^R -
\varphi_\nu^I \varphi_\nu^I) +
4\varphi_\mu^R\varphi_\mu^I\varphi_\nu^R\varphi_\nu^I\right]
\\
+ \frac{i}{\sqrt{8}\Psi_0}(\varphi_\mu^I \partial \varphi_\mu^R -
\varphi_\mu^R \partial \varphi_\mu^I)\cdot \partial\theta +
\frac{1}{16\Psi_0^2}(\varphi_\mu^R \varphi_\mu^R + \varphi_\mu^I
\varphi_\mu^I) (\partial\theta)^2 \bigg\}\punc{.}
\end{multline}
Note that the two fields $\varphi_\mu^R$ and $\varphi_\mu^I$ remain
gapped, but that their gaps are not the same. The lowest-energy
`charged' mode is $\varphi_\mu^I$, with gap $\lambda_I = \sqrt{r -
g^\psi\Psi_0/2}$.

\subsubsection{Self energy}

As an example, we consider the lowest-order diagram that contributes
to the decay rate for $\varphi_\mu^I$, at energies well below that
required to produce an excitation of the field $\varphi_\mu^R$. This
is given by \be \label{SSCSigmaDiagram} \Sigma^\varphi = \;
\parbox{30mm}{
\begin{picture}(30,20)
\put(0,0){\includegraphics{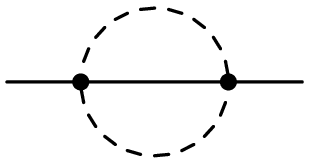}}

\end{picture}}\punc{,}
\ee where the solid lines represent $\varphi_\mu^I$ and the dashed
lines $\theta$. Each vertex represents a factor $\sim (p_1\cdot
p_2)\Psi_0^{-2}$, where $p_1$ and $p_2$ are the momenta of the two
$\theta$ propagators, coming from the final term in $\Act\sub{SSC}$.
(All the other interaction terms involve $\varphi_\mu^R$ and
contribute to the decay rate only at higher energies. As in Section
\ref{SSISelfEnergy}, there is also a lower-order tadpole diagram
that does not contribute to the decay rate.)

\begin{figure*}
\includegraphics{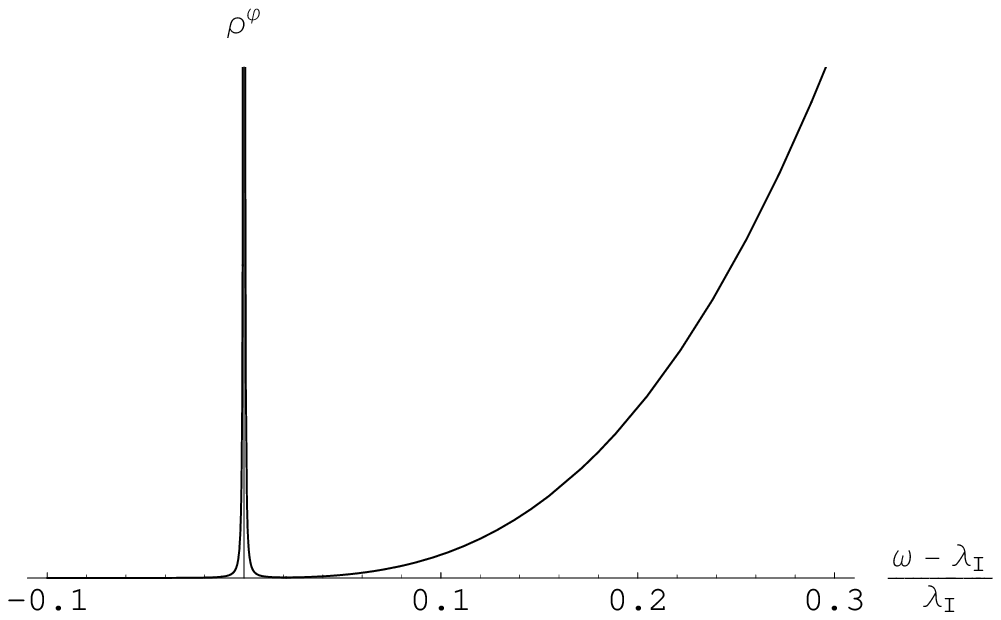}
\caption{\label{fig:SSCrho}The spectral weight $\rho^\varphi$ for
the field $\varphi_\mu^I$ in SSC, near the gap $\lambda_I$,
calculated numerically using the diagram in \refeq{SSCSigmaDiagram}.
As in Figures \ref{fig:SSIrho} and \ref{fig:SSISSCrho}, there is a
delta-function peak at $\omega = \lambda_I$ (which has artificially
been given a nonzero width), corresponding to the stable particle
excitation. In this case, unlike in SSI, the continuum in the
spectral weight occurs immediately above the peak. This is due to
the (gapless) Goldstone mode $\theta$ resulting from the broken
phase symmetry in SSC. The derivatives in the coupling between the
Goldstone mode and the $\varphi_\mu^I$ field in $\Act\sub{SSC}$
strongly suppress the spectral weight as $\omega \rightarrow
\lambda_I$ from above; in fact, $\rho^\varphi \sim
(\omega-\lambda_I)^3$.}
\end{figure*}

The diagram can be calculated numerically and the corresponding
spectral weight is shown in Figure \ref{fig:SSCrho}. As in SSI,
there is a sharp peak (at $\omega = \lambda_I$) corresponding to the
stable gapped `charged' mode, followed at higher energy by a
continuum of excitations. In this case, however, the Goldstone mode
$\theta$ causes the continuum to begin precisely at $\omega =
\lambda_I$, albeit suppressed by a factor of $(\omega-\lambda_I)^3$.

This should be contrasted with the behavior at the transition
itself, described below in Section \ref{SectionRGforSSC}. At the
transition, the gapless modes are critical, rather than Goldstone
modes, and their coupling is not suppressed by powers of the
momentum. As a result, the spectral weight, calculated
perturbatively, does not tend to zero as $\omega \rightarrow
\lambda$ (see Section \ref{SSISSCtransitionPertThy}) and a RG
analysis shows that the sharp quasiparticle peak at $\omega =
\lambda$ is in fact replaced by a weaker power-law singularity,
reflecting the absence of quasiparticles at the critical point.

\subsubsection{Spin response}

The first contribution to the spin response resulting from the
coupling of $\varphi_\mu^I$ to $\theta$ has three loops: \be
\label{Pi1aSSC} \Pi^{(1a)} =\;
\parbox{30mm}{
\begin{picture}(30,20)
\put(0,0){\includegraphics{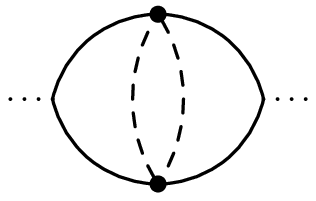}}

\end{picture}
}\punc{.} \ee As for the self energy, the powers of momentum
appearing in the interaction vertex strongly suppress the
contribution at low energy, and this diagram will not modify the
threshold singularity in the spin response.

\subsection{Polar condensate}
\label{PC}

In the polar condensate phase, $\psi_\mu$ condenses, breaking both
the spin- ($S$) and phase-rotation ($\Phi$) symmetries. A different
set of fields is therefore required to describe this phase, in which
there are gapless excitations carrying both particle number and
spin. A similar approach can be used to that described above for
SSC, but since the calculations are rather involved, we will only
give a brief outline.

In terms of the general continuum theory of Section
\ref{sec:ContinuumAction} given in \refeq{Actpsi}, this phase
corresponds to a condensate of $\psi_\mu$. This field is a complex
vector and can take an arbitrary direction in space, which for
simplicity we take as real and along the $z$-axis: $\Mean{\psi_\mu}
= \varphi_0 \delta_{\mu z}$. A convenient representation for the
field is then to write \be \label{DefinePCfields} \psi_\mu =
\varphi_0 \, \e^{\ii \chi}\, \e^{\ii \alpha_i \sigma_i}\, \e^{\ii
\beta_j \tau_j}\, \delta_{\mu z}\punc{.} \ee As in Section
\ref{SSC}, the amplitude modes of the condensed field, which are
inessential to the physics, have been neglected. Summation over
$i,j\in \{x,y\}$ is implied; $\chi$, $\alpha_i$ and $\beta_j$ are
real fields; and $\sigma_i$ and $\tau_j$ are matrices acting in spin
space: $\sigma_x =\left(
\begin{smallmatrix}
0&0&-\ii\\ 0&0&0\\ \ii&0&0
\end{smallmatrix}\right)$,
$\sigma_y =\left(
\begin{smallmatrix}
0&0&0\\ 0&0&-\ii\\ 0&\ii&0
\end{smallmatrix}\right)$,
$\tau_x =\left(
\begin{smallmatrix}
0&0&1\\ 0&0&0\\ 1&0&0
\end{smallmatrix}\right)$ and
$\tau_y =\left(
\begin{smallmatrix}
0&0&0\\ 0&0&1\\ 0&1&0
\end{smallmatrix}\right)$.

The physical interpretation of this parametrization is as follows:
The factor $\e^{\ii \chi}$ incorporates the overall phase degree of
freedom, similarly to $\e^{\ii \theta'}$ in \refeq{SSCphaserep}. The
matrices $\sigma_i$ are the generators of real rotations in spin
space and serve to rotate the axis along which the vector $\psi_\mu$
is aligned. These two kinds of transformation correspond to
symmetries of the action, and the energy is unchanged by uniform
shifts in $\chi$, $\alpha_x$ and $\alpha_y$. There are therefore
three Goldstone modes in this phase, two of which carry spin.

The remaining matrices, $\tau_x$ and $\tau_y$, generate complex
transformations of the condensate wavefunction, and describe the
angular momentum degree of freedom. The corresponding modes,
described by $\beta_x$ and $\beta_y$, would be Goldstone modes in a
system with full $\mathrm{SU}(3)$ symmetry and have gaps
proportional to $v$, the coefficient appearing in \refeq{Actpsi4}.

In this representation, the action can be rewritten \be
\label{PCaction} \Act\sub{PC} = \spacetimeint \left\{ \frac{1}{2}
(\partial\chi)^2 + \frac{1}{2} (\partial \alpha_i)^2 +
\frac{1}{2}\left[(\partial \beta_i)^2 - 4v\beta_i^2\right] -
\frac{2}{\varphi_0} \beta_i (\partial\alpha_i)\cdot(\partial\chi) +
\cdots \right\}\punc{,} \ee where summation over $i$ is again
implied. [As in \refeq{SSCactionPsi}, the fields have been rescaled
by constant factors to give the coefficients of the kinetic terms
their conventional values.]

The angular momentum, given by \refeq{L:field}, can be rewritten in
terms of the fields defined in \refeq{DefinePCfields}, giving for
the transverse components \be \label{PCTransverseL} L_x \sim
-\beta_y\qquad\mathrm{and}\qquad L_y \sim \beta_x\punc{,} \ee and
for the longitudinal component \be L_z \sim \alpha_x \beta_y -
\alpha_y \beta_x\punc{.} \ee The longitudinal and transverse
components are given by different expressions because of the broken
spin-rotation symmetry. These different expressions will lead to
qualitatively different forms for the transverse and longitudinal
spin responses.

\section{Quantum phase transitions}
\label{sec:qpt}

We now describe the behavior of the system across transitions out of
the spin-singlet insulator (SSI) into the various phases described
above. First we briefly address each of the transitions in turn,
before describing them in more detail, in Sections
\ref{SectionRGforPC}, \ref{SectionRGforSSC} and \ref{SectionRGd}.

First, consider the transition from SSI into the nematic insulator
(NI). Both phases have $\Mean{\psi_\mu} = 0$, and off-diagonal
elements of $\Mean{\bar{\psi}_\mu\psi_\nu}$ become nonzero across
the transition. As suggested by the mean-field analysis Section
\ref{SectionQuantumRotors}, this is expected to be a first-order
transition. This is confirmed by the presence of terms cubic in
$\bar{\psi}_\mu\psi_\nu$ in the action $\Act_\psi$, \refeq{Actpsi},
which are not forbidden by any symmetry (and are hidden in the
ellipsis). We will not consider this transition further here.

\subsection{SSI to PC}
\label{SSItoPC}

At the transition to PC, the field $\psi_\mu$ becomes critical; the
appropriate field theory is therefore given by \refeq{SSIaction}.
Since the critical field $\psi_\mu$ carries spin, this transition is
an example of class {\bf A} identified above, in Section
\ref{sec:intro}. The field $\psi_\mu$ is gapless and the spin
response will be governed by the correlators of
$\bar\psi_\mu\psi_\nu$, as in SSI. A RG analysis of this transition
will be presented below in Section \ref{SectionRGforPC}.

\subsection{SSI to SSC}
\label{SSItoSSC}

The critical field at the transition to SSC is the singlet pair
$\Psi$, introduced in Section \ref{SSInearSSC}. Once $\Psi$ has been
isolated, $\psi_\mu$, which has no gapless excitations, can be
safely integrated out. This leaves the field theory of a single
complex scalar, with the same form as the action $\Act_\Psi$ given
in \refeq{ActPsiSSI}: \be \label{ActPsi} \Act_\Psi =
\spacetimeint\left(|\partial \Psi|^2 + r_\Psi |\Psi|^2 +
\frac{u_\Psi}{4}|\Psi|^4 + \,\cdots\right)\punc{.} \ee This
transition is therefore of the XY universality class, with upper
critical dimension $D = d + 1 = 4$. (The field $\Psi$ has
engineering dimension $[\Psi] = (d-1)/2$, so the coupling $u_\Psi$
has dimension $[u_\Psi] = 3-d$.)

Since the critical field $\Psi$ is spinless, this phase transition
falls within class {\bf B} identified in Section \ref{sec:intro}.
While the action $\Act_\Psi$ is sufficient to describe the critical
properties of the ground state across the transition, we are
primarily interested in excitations that carry spin, and the
critical theory given by $\Act_\Psi$ does not describe these.
Instead, we must keep the singly-charged excitations given by
$\psi_\mu$, and use the full action $\Act\sub{SSI}'$, in
\refeq{SSInearSSCaction}.

As in Section \ref{SSInearSSC}, the spin response is determined by
the lowest-energy spin-carrying (but overall charge-neutral)
excitations. These will normally be described by particle--hole
pairs of $\psi_\mu$, but we also address the case in which there is
a lower-energy bound state, which then governs the long-time
response. Such a bound state is more likely to form in a region of
the SSI-to-SSC phase boundary which is well away from the PC phase
in Fig.~\ref{MeanFieldPhaseDiagram}.

\subsubsection{Without bound state}
\label{SectionRegionC}

When there is no bound state, the spin response is determined by the
correlators of the compound operator $\bar{\psi}_\mu \psi_\nu$, as
in \refeq{PiFromPsi}. The corresponding action is therefore given by
$\Act\sub{SSI}'$, in \refeq{SSInearSSCaction}.

Since $\psi_\mu$ has only gapped excitations, while the field $\Psi$
is now gapless, this can be simplified somewhat. The important
excitations are those just above the gap $\lambda = \sqrt{r}$, for
which the dispersion can be replaced by a nonrelativistic form. We
define particle and hole operators so that $\psi_\mu \sim p_\mu +
\bar{h}_\mu$, giving an action $\Act_\Psi + \Act_{\Psi,\psi}'$,
where
\begin{multline}
\Act'_{\Psi,\psi} = \spacetimeint \bigg[ \bar{p}_\mu \left(\ii\partial_t - \frac{\nabla^2}{2m_\psi} + \lambda\right) p_\mu + \bar{h}_\mu \left(\ii\partial_t - \frac{\nabla^2}{2m_\psi} + \lambda\right) h_\mu\\
+ g_\psi \left( \bar\Psi p_\mu \bar{h}_\mu + \Psi \bar{p}_\mu h_\mu
\right)\bigg]\punc{.}
\end{multline}

Using power counting (and taking $[t] = [\xv] = -1$, since the
critical theory $\Act_\Psi$ is relativistic), the engineering
dimension of the kinetic-energy term is $[1/m_\psi] = -1$. The
dispersion is therefore irrelevant and, at least \primafacie, the
particles and holes can be treated as static impurities. We
therefore take for the action \be \Act_{\Psi,\psi} =
\spacetimeint\left[\bar{p}_\mu \left(\ii\partial_t + \lambda\right)
p_\mu + \bar{h}_\mu \left(\ii\partial_t + \lambda\right) h_\mu +
g_\psi \left( \bar\Psi p_\mu \bar{h}_\mu + \Psi \bar{p}_\mu h_\mu
\right)\right]\punc{.} \ee

The scaling dimension of the coupling $g_\psi$ is $[g_\psi] =
(3-d)/2$, so that it is relevant for $d < 3$. It is therefore
relevant in two (spatial) dimensions and marginal in three, and we
will consider both of these cases below. Any other interactions,
including terms quartic in $p_\mu$ and $h_\mu$, are irrelevant.

In Section \ref{SectionRGforSSC}, we will treat the case $d=2$ using
a renormalization-group (RG) analysis, and then, in Section
\ref{SSISSCtransitionPertThy}, return to the case $d=3$, where
straightforward perturbation theory is sufficient.

\subsubsection{With bound state}
\label{SectionRegionD}

If a bound state of $p_\mu$ and $h_\nu$ exists below the continuum,
it will determine the lowest-energy spin response. At higher
energies, the continuum of particle and hole excitations remains and
will have the same effects as described above in Section
\ref{SectionRegionC}.

The particle and hole excitations $p_\mu$ and $h_\nu$ carry spin
$1$, so the simplest spin-carrying bound states are a quintet with
spin $2$. (Bound states with antisymmetric relative spatial
wavefunctions and net spin $S = 1$ are also possible.) To describe
these bound states, we introduce the field $d_{\mu\nu}$, a
(symmetric, traceless) second-rank tensor: \be d_{\mu\nu} \sim
\bar\psi_\mu\psi_\nu - \frac{1}{3}\delta_{\mu\nu}\bar\psi_\rho
\psi_\rho\punc{.} \ee

By symmetry, angular momentum $\Lv$ couples to $d_{\mu\nu}$ by
$L_\rho \sim
\ii\epsilon_{\mu\nu\rho}\bar{d}_{\mu\lambda}d_{\lambda\nu}$, and it
is necessary to create a pair of $d$ excitations to propagate the
spin excitation. This will require more energy than an unbound
particle-hole pair and so the response function $\Pi_\Lv$ will still
be given by the compound operator $\bar\psi_\mu\psi_\nu$, as in
Section \ref{SectionRegionC}.

On the other hand, it is clear from its definition, in
\refeq{DefineQ}, that $Q_{\mu\nu}$ couples directly to $d_{\mu\nu}$.
The response function $\Pi_Q$ is therefore given by the two-point
correlator of $d_{\mu\nu}$: \be \label{PiQfromd} \Pi_Q \sim
\Mean{\mathcal{T}_t\: \bar{d}_{\mu\nu}(\xv,t)
d_{\mu\nu}(\zerov,0)}\punc{.} \ee

To construct the corresponding action, we notice first that there is
no term such as $\bar\Psi \Psi d_{\mu\mu}$, because $d_{\mu\nu}$ is
traceless, so that the next order in the expansion must be taken.
The action for $d_{\mu\nu}$ is then given by \be \Act_{\Psi,d} =
\spacetimeint\left[\bar{d}_{\mu\nu}\left(\ii\partial_t -
\frac{\nabla^2}{2m_d} + r_d \right) d_{\mu\nu} + g_d
\bar{d}_{\mu\nu}d_{\mu\nu}\bar\Psi \Psi +\, \cdots\right]\punc{.}
\ee

The same power-counting argument as above shows that $[1/m_d] = -1$
and the dispersion is irrelevant. The scaling dimension of the
interaction $g_d$ is $[g_d] = 2 - d$, so that it becomes marginal at
$d=2$ and is irrelevant for $d=3$. This case therefore falls within
class {\bf B2}; we describe the properties of the critical region
below, in Section \ref{SectionRGd}.

\section{Critical properties: SSI/PC (Class A)}
\label{SectionRGforPC}

The transition from SSI to PC is described by the action given in
\refeq{Actpsi}, with two different quartic terms allowed by the
$\mathrm{U}(1)\otimes \mathrm{O}(3)$ symmetry: \be \Act_\psi =
-\bar{\psi}_\mu \partial^2 \psi_\mu +  r\bar{\psi}_\mu \psi_\mu +
\frac{u}{4}\bar\psi_\mu \psi_\mu \bar\psi_\nu \psi_\nu +
\frac{v}{4}\bar\psi_\mu\bar\psi_\mu\psi_\nu\psi_\nu\punc{.} \ee

To study the critical behavior of the model, a straightforward RG
calculation can be performed, leading to the following beta
functions to one loop:
\begin{align}
\beta_u &= -\epsilon \tilde u + \frac{7}{2}\tilde u^2 + 2\tilde u\tilde v + 2\tilde v^2\\
\beta_v &= -\epsilon \tilde v + 3 \tilde u\tilde v +
\frac{3}{2}\tilde v^2
\end{align}
(where $\tilde u$ and $\tilde v$ are related to $u$ and $v$ by
constant factors). These flow equations have no stable fixed point
for finite $u$ and $v$. However, far more complete six-loop analyses
in \explicitcite{PratoPelissetto} have shown that a stable fixed
point does indeed exist with $v<0$. At this fixed point, the full
Green function for the field $\psi_\mu$ behaves like \be G^\psi(p)
\sim p^{-2+\eta_\psi}\punc{,} \ee where $\eta_\psi = 0$ to one-loop
order, while the six-loop estimate is \cite{PratoPelissetto}
$\eta_\psi \approx 0.08$. The polarization $\Pi_{\mu\nu,\rho\sigma}$
is determined by the response functions $\Pi_\Lv$ and $\Pi_Q$
defined in Section \ref{sec:PropertiesOfPhases}, and these have
singularities of the form \be \Pi_\Lv(q) \sim
q^{-2+\eta_\Lv}\punc{,}~~\Pi_Q(q) \sim q^{-2+\eta_Q}\punc{,} \ee
where the $\eta_{\Lv,Q}$ are exponents related to scaling dimensions
of operators bilinear in $\psi_\mu$ at the fixed point of
\explicitcite{PratoPelissetto}. In particular, we have $\eta_\Lv =
\eta_H$ where the latter exponent is defined in
\explicitcite{PratoPelissetto}, and for which their estimate is
$\eta_H \approx 2.70$. For $\Pi_Q$ we have $\eta_Q = d+3 -2y_3$,
where $y_3 \approx 2.0$ is the exponent listed in Table III of
\explicitcite{CalabreseVicari} for the collinear case with $N=3$.

\section{Critical properties: SSI--SSC (Class B1)}
\label{SectionRGforSSC}

The critical theory for the phase transition between SSI and SSC is
given by \be \label{PsiCriticalTheory} \Act_\Psi =
\spaceitimeint\left(|\partial \Psi|^2 + r_\Psi |\Psi|^2 +
\frac{u_\Psi}{4}|\Psi|^4\right)\punc{,} \ee where we have rewritten
the action in imaginary time. Since the field $\Psi$ carries no
spin, the lowest energy excitations are described by \be
\Act_{\Psi,\psi} = \spaceitimeint\left[\bar{p}_\mu
\left(\partial_\tau + \lambda\right) p_\mu + \bar{h}_\mu
\left(\partial_\tau + \lambda\right) h_\mu + g_\psi \left( \bar\Psi
p_\mu \bar{h}_\mu + \Psi \bar{p}_\mu h_\mu \right)\right]\punc{.}
\ee

For $d<3$, the correlation functions of the particle and hole
excitations can be found using a RG calculation. Since the present
approach is slightly different from the standard RG, we perform the
calculation using a cutoff in momentum space, which makes the logic
involved more transparent, in Appendix \ref{CutoffRG}. Here we use
dimensional regularization, which is the simplest approach from a
calculational point of view.

We define the (imaginary-time) free propagator for the $\Psi$ field
as \be G^\Psi_0(\kv,\ii\omega) = \frac{1}{k^2+\omega^2 +
r_\Psi}\punc{.} \ee At the critical point, the renormalized mass of
$\Psi$ vanishes; in dimensional regularization, this occurs for
$r_\Psi = 0$. For the $p_\mu$ and $h_\mu$ fields, the propagator is
\be G^\psi_0(\ii\omega) = \frac{1}{-\ii\omega + \lambda}\punc{,} \ee
independent of the momentum.

The renormalization of the terms in the action $\Act_\Psi$
describing $\Psi$ is identical to the standard
analysis:\cite{Zinn-Justin} the presence of the gapped $\psi_\mu$
excitations cannot affect the critical behavior of the gapless
$\Psi$ field. The corresponding RG has a fixed point with $u_\Psi$
of order $\epsilon = 3-d$, at which the scaling dimension of $\Psi$
is $[\Psi] = 1 - \epsilon/2 + \epsilon^2/100 + \Order{\epsilon^3}$.

To lowest order in the coupling $g_\psi$ (or, as will subsequently
be shown to be equivalent, in an expansion in $\epsilon$), the only
self-energy diagram for the particle field $p_\mu$ is as shown in
\refeq{ParticleHolePairBubble}:
\begin{align}
\Sigma^\psi_1(\ii\omega) &=\qquad
\parbox{30mm}{
\begin{picture}(30,20)
\put(0,0){\includegraphics{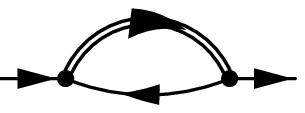}}
\fmfL(15.00029,17.94228,b){$\kv ,\ii \omega '$}%

\end{picture}}
\label{SelfEnergyDiagram1}
\\
&= g_\psi^2\int_\kv \int_{-\infty}^{\infty}\frac{\dd\omega'}{2\pi}
\, G^\Psi_0(\kv,\ii\omega')
G^\psi_0\biglb(\ii(\omega-\omega')\bigrb)\punc{,}
\end{align}
where \be \int_\kv \equiv \int \frac{\dd^d \kv}{(2\pi)^d} \equiv
\Omega_d \int_0^\infty\dd k\, k^{d-1} \ee (for an isotropic
integrand).

There happen to be no diagrams giving a renormalization of the
coupling $g_\psi$ at this order, but such diagrams appear in higher
orders, as in \refeq{2loopG}.

Charge-neutral compound operators such as $L_\rho$ and $Q_{\mu\nu}$,
defined in Section \ref{sec:Observables}, can be written in the form
$T_{\mu\nu} p_\mu h_\nu$, where $T_{\mu\nu}$ is a matrix of
$c$-numbers. The critical exponents for these operators can then be
found by considering the renormalization of the corresponding
insertion, given by
\begin{align}
H_1(2\ii\omega) &=\qquad
\parbox{25mm}{
\begin{picture}(25,25)
\put(0,0){\includegraphics{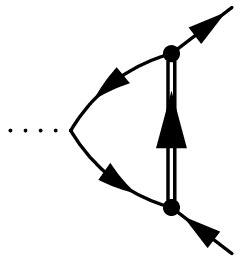}}
\fmfL(20.99109,3.80617,lb){$\mu $}%
\fmfL(9.18875,5.8954,rt){$\mu $}%
\fmfL(18.47226,12.49998,l){$\kv ,\ii \omega '$}%
\fmfL(9.18878,19.10498,rb){$\nu $}%
\fmfL(20.99109,21.19383,lt){$\nu $}%

\end{picture}}
\label{InsertionDiagram1}
\\
&= g_\psi^2 \int_\kv \int_{-\infty}^{\infty}\frac{\dd\omega'}{2\pi}
\, G^\Psi_0(\kv,\ii\omega')
G^\psi_0\biglb(\ii(\omega-\omega')\bigrb)
G^\psi_0\biglb(\ii(\omega+\omega')\bigrb)\punc{,}
\end{align}
where all spin indices have been omitted in the latter expression.
In accounting for these indices, it is important to note that the
exchange of the pair interchanges the particle and hole lines and
hence $\mu$ and $\nu$. This causes the results to be dependent on
the symmetry of the matrix $T_{\mu\nu}$, leading to different
scaling exponents for excitations with even ($T_{\mu\nu}$ symmetric,
eg, $Q_{\mu\nu}$) and odd spin ($T_{\mu\nu}$ antisymmetric, eg,
$L_\rho$).

To order $\epsilon^2$, the diagrams that must be evaluated are, for
the self energy: \be \Sigma^\psi_2(\ii\omega) =\quad
\parbox{35mm}{
\begin{picture}(35,20)
\put(0,0){\includegraphics{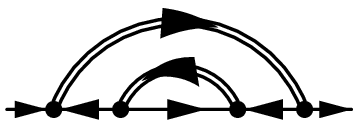}}

\end{picture}}
\ee and for the renormalization of $g_\psi$: \be \label{2loopG}
\Gamma^\psi_2(\ii\omega) =\quad
\parbox{35mm}{
\begin{picture}(35,25)
\put(0,0){\includegraphics{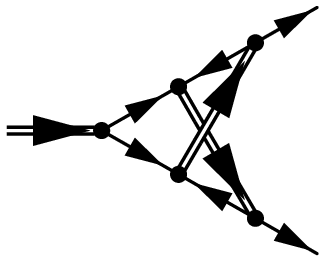}}

\end{picture}}
+\quad
\parbox{35mm}{
\begin{picture}(35,25)
\put(0,0){\includegraphics{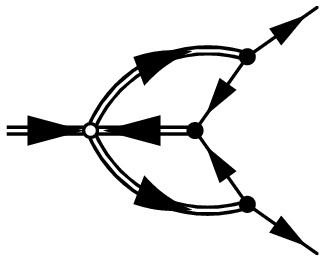}}

\end{picture}}
\ee (Note that the second diagram involves the four-point coupling
of the $\Psi$ field, indicated in the diagram by the empty circle.)
The renormalization of the insertion is given by the following
diagrams: \be \label{2loopH} H_2(2\ii \omega) = \quad
\parbox{35mm}{
\begin{picture}(35,25)
\put(0,0){\includegraphics{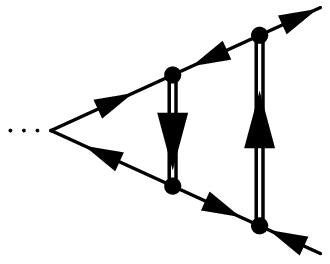}}

\end{picture}}
+\quad
\parbox{35mm}{
\begin{picture}(35,25)
\put(0,0){\includegraphics{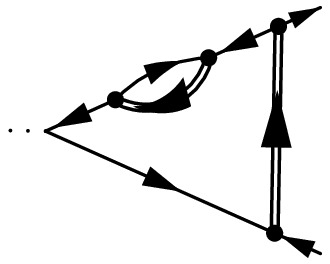}}

\end{picture}}
+\quad
\parbox{35mm}{
\begin{picture}(35,25)
\put(0,0){\includegraphics{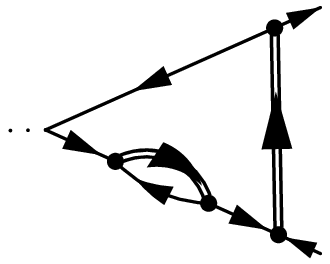}}

\end{picture}}
\ee

We will display the steps in the calculation of the results to order
$\epsilon$ and simply state the higher-order results afterwards. For
the self energy, performing the integral over $\omega'$ using
contour integration gives \be \Sigma^\psi_1(\ii\omega) =
g_\psi^2\int_\kv \frac{1}{2k}\cdot\frac{1}{-\ii\omega + \lambda +
k}\punc{,} \ee which leads to, defining $z = -\ii\omega + \lambda$,
\be \Sigma^\psi_1(\ii\omega) = -\frac{g_\psi^2
z^{1-\epsilon}}{4\pi^2\epsilon}\punc{,} \ee plus terms that are
finite as $\epsilon\rightarrow 0$.

To this order, the full propagator of the particle is then given by
\begin{align}
(G^\psi)^{-1} &= z - \Sigma^\psi_1\\
&= z\left(1 + \frac{g_\psi^2
z^{-\epsilon}}{4\pi^2\epsilon}\right)\punc{,}
\end{align}
so that renormalizing the propagator (using minimal subtraction) at
frequency $z = \mu$ gives for the wavefunction renormalization \be
Z_\psi = 1 - \frac{g_\psi^2 \mu^{-\epsilon}}{4\pi^2\epsilon}\punc{.}
\ee Since there are no diagrams corresponding to renormalization of
the coupling, we have $Z_g = 1$, to this order.

We now define the (dimensionless) renormalized coupling $\tilde
g_\psi$, given by \be \label{Definegtilde} g_\psi = 2\pi \tilde
g_\psi \frac{\mu^{\epsilon/2}Z_g}{Z_\psi\sqrt{Z_\Psi}}\punc{.} \ee
In terms of this, we have $Z_\psi = 1 - {\tilde g_\psi}^2/\epsilon$.
The beta function for the coupling is then given by \be
\label{BetaFunction} \beta(\tilde g_\psi) \equiv \mu
\parderat{\tilde g_\psi}{\mu}{g_\psi} =
\tilde{g}_\psi\left(-\frac{\epsilon}{2} + {{\tilde
g}_\psi}^2\right)\punc{,} \ee so that the fixed point is at \be
\label{FixedPoint} \tilde{g}_\psi^\star =
\sqrt{\frac{\epsilon}{2}}\punc{.} \ee Since the fixed point has
$\tilde{g}_\psi^\star \sim \epsilon^{1/2}$, a perturbative expansion
at this point is indeed equivalent to an expansion in $\epsilon$.

The anomalous dimension of the particle (and hole) propagator is
then given by \be \eta_\psi = \beta \frac{\dd}{\dd
\tilde{g}_\psi}\log Z_\psi\punc{,} \ee so that, to first order,
$\eta_\psi = \epsilon/2$ at the fixed point.

The two-point Green function behaves, for $\omega > \lambda$, as \be
\label{TwoPointCorrelatorResult} G^\psi(\omega) \sim (\lambda -
\omega)^{-1+\eta_\psi}\punc{,} \ee so that the corresponding
spectral weight is given by \be \rho^\psi(\omega) \sim (\omega -
\lambda)^{-1+\eta_\psi}\punc{.} \ee The relativistic invariance of
the original theory allows these results to be extended to finite
external momentum by the usual replacement $\omega \rightarrow
\sqrt{\omega^2 - k^2}$.

\begin{figure}
\begin{center}
\includegraphics{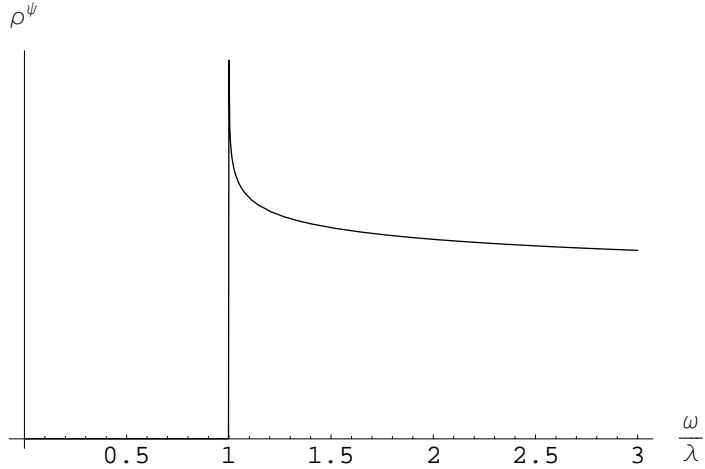}
\end{center}
\caption{\label{fig:SSISSCrho2}The spectral weight $\rho^\psi$ at
the SSI--SSC transition in $d=2$ spatial dimensions. The
delta-function peak at $\omega = \lambda$ (present in the SSI and
SSC phases) has been replaced by a continuum of excitations, with
$\rho^\psi \sim (\omega - \lambda)^{-1+\eta_\psi}$. The numerical
value $\eta_\psi = 0.91797$ used in the plot results from a
dimensional expansion in $\epsilon = 3-d$, carried out to order
$\epsilon^2$, \refeq{AnomalousDimension}, and evaluated at $\epsilon
= 1$.}
\end{figure}

The results at the next order in this expansion also involve
diagrams renormalizing the coupling $g_\psi$. The fixed point then
occurs at \be (\tilde{g}_\psi^\star)^2 = \frac{1}{2}\epsilon -
\left(\frac{\pi^2}{15}-\frac{49}{100}\right)\epsilon^2 +
\Order{\epsilon^3}\punc{,} \ee and the anomalous dimension is \be
\label{AnomalousDimension} \eta_\psi = \frac{1}{2}\epsilon +
\left(\frac{\pi^2}{15}-\frac{6}{25}\right)\epsilon^2 +
\Order{\epsilon^3}\punc{.} \ee Figure \ref{fig:SSISSCrho2} shows the
spectral weight $\rho^\psi$ for $d=2$ ($\epsilon = 1$), using the
numerical value from \refeq{AnomalousDimension}. The quasiparticle
peak appearing on both sides of the critical points, Figures
\ref{fig:SSIrho}, \ref{fig:SSISSCrho} and \ref{fig:SSCrho}, is
replaced by an incoherent continuum of excitations.

The scaling exponents for compound operators of the form $T_{\mu\nu}
p_\mu h_\nu$ can be found in the same way, and the corresponding
anomalous exponents are given by \be \eta_{\Pi_\pm} = \beta
\frac{\dd}{\dd \tilde{g}_\psi}\log
\frac{Z_{\Pi_\pm}}{Z_\psi}\punc{,} \ee for symmetric ($+$) and
antisymmetric ($-$) matrices $T_{\mu\nu}$. The correlators of these
operators, $\Pi_\pm^L$, then yield the scaling dimensions of
observables closely related to the $\Pi$ spin correlators defined in
Section~\ref{sec:SSIspinresponse}. This relationship will be
discussed more explicitly in Section~\ref{SectionHeuristic} below,
where we will see that it is necessary to include the formally
irrelevant momentum dispersion of the $p_\mu$ and $h_\mu$ particles
to obtain the low momentum spin response functions of
Section~\ref{sec:SSIspinresponse}. From this argument we will obtain
the general result that
\begin{equation}
\mbox{dim}[\Pi] = \mbox{dim}[\Pi^L] + d/2 \punc{,} \label{ss3}
\end{equation}
where the corresponding $\Pi$ observables are taken on both sides of
the equation. The correlation function is then given, for $\omega >
\lambda$, by \be \label{PiPowerLaw} \Pi_\pm^L (2\omega) \sim
(\lambda - \omega)^{-1+\eta_{\Pi_\pm}}\punc{.} \ee and the
corresponding spectral density, \be A_\pm^L (2\omega) =
\lim_{\varepsilon\to 0^+}\im\Pi_\pm^L (2\omega +
\ii\varepsilon)\punc{,} \ee is a delta-function at $\omega =
\lambda$ for $\eta_\Pi = 0$, and otherwise behaves like \be
\label{APowerLaw} A_\pm(2\omega)^L \sim (\omega -
\lambda)^{-1+\eta_{\Pi_\pm}}\punc{,} \ee for $\omega$ just above the
gap $\lambda$. The physical spin correlation therefore has the
spectral density, from Eq.~(\ref{ss3}),
\begin{equation}
A_\pm(2\omega) \sim (\omega -
\lambda)^{(d-2)/2+\eta_{\Pi_\pm}}\punc{.} \label{ss4}
\end{equation}
This result should be compared with the spectral density of
Eq.~(\ref{ss1}), and the discussion below it, which holds both in
the SSI and SSC phases.

 The
perturbation calculation including diagrams up to two loops gives
the exponent \be \eta_{\Pi_+} = - \epsilon + \left(
\frac{2\pi^2}{15} - \frac{49}{50}\right) \epsilon^2 +
\Order{\epsilon^3}\punc{,}  \label{ss5} \ee for operators with even
spin, such as $Q_{\mu\nu}$, and $\eta_{\Pi_-} = 0$ exactly for those
with odd spin, such as the angular momentum $L_\rho$. The latter
result is a consequence of the conservation of angular momentum.

\subsection{Effects of dispersion}

As argued in Section \ref{SSItoSSC}, the dispersion of the particle
and hole are formally irrelevant in the RG calculation. Here we will
present a few more details of the argument, and also show how the
singularity in the physical momentum conserving spin response
functions $\Pi$ in Section~\ref{sec:SSIspinresponse} are related to
the local observables $\Pi^L$ by Eq.~(\ref{ss3}).

\subsubsection{Scaling form}
\label{ScalingForm}

We first describe the standard scaling argument that suggests
dispersion can be ignored. We concentrate here on the function
$A_{\mu\nu,\rho\sigma}$, which is sufficient to determine
$\Pi_{\mu\nu,\rho\sigma}$ via the Kramers-Kronig relations. (Similar
considerations to the following apply to $G^\psi$.)

Consider the action of the RG transformation on the correlation
function, exactly at the critical point. Let $\alpha =
\frac{1}{2m_\psi}$ be the coefficient of $k^2$ in the quadratic part
of the action of $p_\mu$ and $h_\mu$ and suppose that this is small
(in some sense) but nonzero. Using real frequencies, with the
shorthand $\delta\omega = \omega - \lambda$, and suppressing the
spin indices, scaling implies \be \label{ScalingOfPiWithDispersion1}
A\biglb(2(\lambda + \delta\omega), \alpha\bigrb) = b^{x_\Pi}A\biglb(
2(\lambda + b\delta\omega),b^{y_\alpha}\alpha\bigrb)\punc{,} \ee
where $x_\Pi = 1 - \eta_\Pi$ and $y_\alpha$ is the scaling dimension
of the operator corresponding to dispersion. By dimensional
analysis, $y_\alpha = -1 + \Order{\epsilon}$, so the dispersion is
irrelevant sufficiently close to dimension $d = 3$ including, we
assume, at the physically important case of $d = 2$, $\epsilon = 1$.

Using \refeq{ScalingOfPiWithDispersion1} we can write the scaling
form \be A\biglb(2(\lambda + \delta\omega), \alpha\bigrb) \sim
\delta\omega^{-x_\Pi}
\mathcal{A}(\delta\omega^{-y_\alpha}\alpha)\punc{.} \ee The power
law in \refeq{APowerLaw} corresponds to taking the limit $\alpha
\rightarrow 0$ in this expression, and---assuming analyticity at
this point---is therefore valid when $\delta\omega^{-y_\alpha}\alpha
\ll 1$. Since $y_\alpha < 0$, we require that $\delta\omega$ be
sufficiently small. In other words, the dispersionless result for
$A(2\omega)$ is appropriate for $\omega$ just larger than $\lambda$.

\subsubsection{Perturbation theory}
\label{SectionHeuristic}

To provide a heuristic guess at the modifications caused by
dispersion, we first consider the two-point correlator for the
fields at finite momentum, $G^\psi(\kv, \omega)$. The simplest
effect that can be expected is the replacement of the gap $\lambda$
by a momentum-dependent energy $\lambda + \alpha k^2$, corresponding
to the free dispersion.

This gives the slightly modified result \be G^\psi(\kv,\omega) \sim
\left(\lambda + \alpha k^2 - \omega\right)^{-1+\eta_\psi}\punc{.}
\ee For fixed $\kv$, there remains a single peak, shifted from its
$\kv = 0$ position by a relatively small amount.

To estimate the effect of dispersion on $\Pi_{\mu\nu,\rho\sigma}$,
consider the lowest-order diagram, evaluated at external momentum
equal to zero:
\begin{align}
\Pi^{(0)}(2\ii\omega) &= \qquad
\parbox{30mm}{
\begin{picture}(30,20)
\put(0,0){\includegraphics{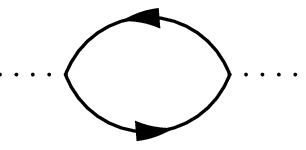}}

\end{picture}
}
\\
&=\int_{\kv,\omega'} G^\psi_0\biglb(\kv,\ii(\omega + \omega')\bigrb)G^\psi_0\biglb(-\kv,\ii(\omega-\omega')\bigrb)\\
&=\frac{1}{2}\int_\kv \frac{1}{\lambda + \alpha{k^2} - \ii\omega}
\punc{.} \label{Pi0Expression}
\end{align}
Note that \refeq{Pi0Expression} corresponds to replacing the
dispersionless result, $\sim(\lambda - \ii\omega)^{-1}$, by $-i
\omega \rightarrow -i \omega + \alpha k^2$, and summing over all
momenta in the loop. It is clear that the same replacement can be
made also in other higher order corrections with $k$ replaced by
other small momenta carried by the $p_\mu$ or $h_\mu$ quanta. We can
heuristically account for such corrections in the above expression
by writing (in real frequency)
\begin{align}
\Pi(2\omega) &\sim \int_\kv\left(\lambda + \alpha{k^2} - \omega\right)^{-1+\eta_{\Pi}}\\
&\sim \int_0^\Lambda \dd k\, k^{d-1} \left(\lambda + \alpha{k^2} -
\omega\right)^{-1+\eta_{\Pi}} \label{DispersionCorrectionIntegral}
\punc{,}
\end{align}
where a momentum cutoff $\Lambda$ has been used. After taking the
imaginary part, the integral can be performed to yield
\begin{equation}
A (2 \omega) \sim \left\{
\begin{array}{cc} (\omega-\lambda)^{(d-2)/2 + \eta_\Pi} &
~~\mbox{for $\lambda < \omega \ll \Delta$} \\
(\omega-\lambda)^{-1+ \eta_\Pi} & ~~\mbox{for $ \omega \gg \Delta$}
\end{array} \right.
\end{equation}
where $\Delta \sim \alpha \Lambda^2$ is the bandwidth of the
$p_\mu$, $h_\mu$ excitations. Notice that the answer associated with
the local spectral density, $A^L$, appears at frequencies larger
than the bandwidth, while the threshold singularity obeys
Eq.~(\ref{ss3}).

\subsection{At and above the upper critical dimension}
\label{SSISSCtransitionPertThy}

The critical results so far have been for $d<3$, where they are
controlled by non-zero fixed point value of $\tilde{g}_\psi^\star$.

For $d = 3$, the fixed point of the RG equations occurs for
$\tilde{g}_\psi^\star = 0$, so perturbation theory in the coupling
can be used to determine the structure of the Green function. Using
the fully relativistic form of the action to calculate the
lowest-order self-energy diagram gives, for $\omega^2 > \lambda^2 +
k^2$, \be \im \Sigma^\psi_{d=3} = \frac{g_\psi^2}{8\pi}\cdot
\frac{\omega^2 - k^2 - \lambda^2}{\omega^2 - k^2}\punc{.} \ee For
comparison, the same calculation in two dimensions gives \be \im
\Sigma^\psi_{d=2} = \frac{g_\psi^2}{8}\cdot \frac{1}{\sqrt{\omega^2
- k^2}}\punc{.} \ee ($\im \Sigma^\psi = 0$ for $\omega^2 < \lambda^2
+ k^2$ in both cases.)

\begin{figure}
\begin{center}
\includegraphics{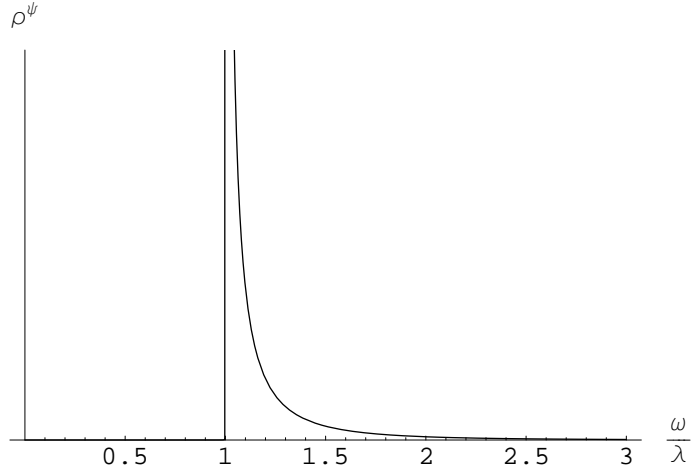}
\end{center}
\caption{\label{fig:SSISSCrho3}The spectral weight $\rho^\psi$ at
the SSI--SSC transition in $d=3$ spatial dimensions. The
delta-function peak at $\omega = \lambda$ has been replaced by a
continuum of excitations, with the mean-field exponent $\rho^\psi
\sim (\omega - \lambda)^{-1}$.}
\end{figure}

Note that $\im \Sigma^\psi_{d=2}$ tends to a constant as $\omega
\rightarrow \sqrt{\lambda^2 + k^2}$ from above, in contrast to the
cases considered in Section \ref{sec:PropertiesOfPhases} above. The
same is not true in three dimensions, and we have \be \im
\Sigma^\psi_{d=3} \sim \omega - \sqrt{\lambda^2 + k^2}\punc{,} \ee
and in general the threshold singularity is of the form $(\omega -
\lambda)^{(d-2)}$. For $d<3$, this leading order estimate of the
threshold singular is not correct, and it is necessary to resum
higher order contribution by the RG, and was done in the previous
subsection. For $d>3$, higher order corrections are subdominant, and
the leading order result here yields the correct singularity.
Finally for $d=3$, we expect from Eq.~(\ref{BetaFunction}) that the
coupling constant will acquire a logarithmic frequency dependence,
$g_\psi^2 \sim 1/\log(\omega-\lambda)$, and this will modify the
above results by a logarithmic prefactor. The spectral weight for
the particle and hole excitations in $d=3$ (without the higher order
logarithmic correction) is shown in Figure \ref{fig:SSISSCrho3}. As
in Figure \ref{fig:SSISSCrho2}, the coherent quasiparticle peak is
replaced by a continuum of excitations, but the exponent is given by
its mean-field value: $\rho^\psi \sim (\omega - \lambda)^{-1}$.
Note, however, that once the logarithmic correction has been
included, the quasiparticle peak remains marginally stable in $d=3$.
The quasiparticle peak is well-defined for $d>3$.

\begin{figure*}
\includegraphics{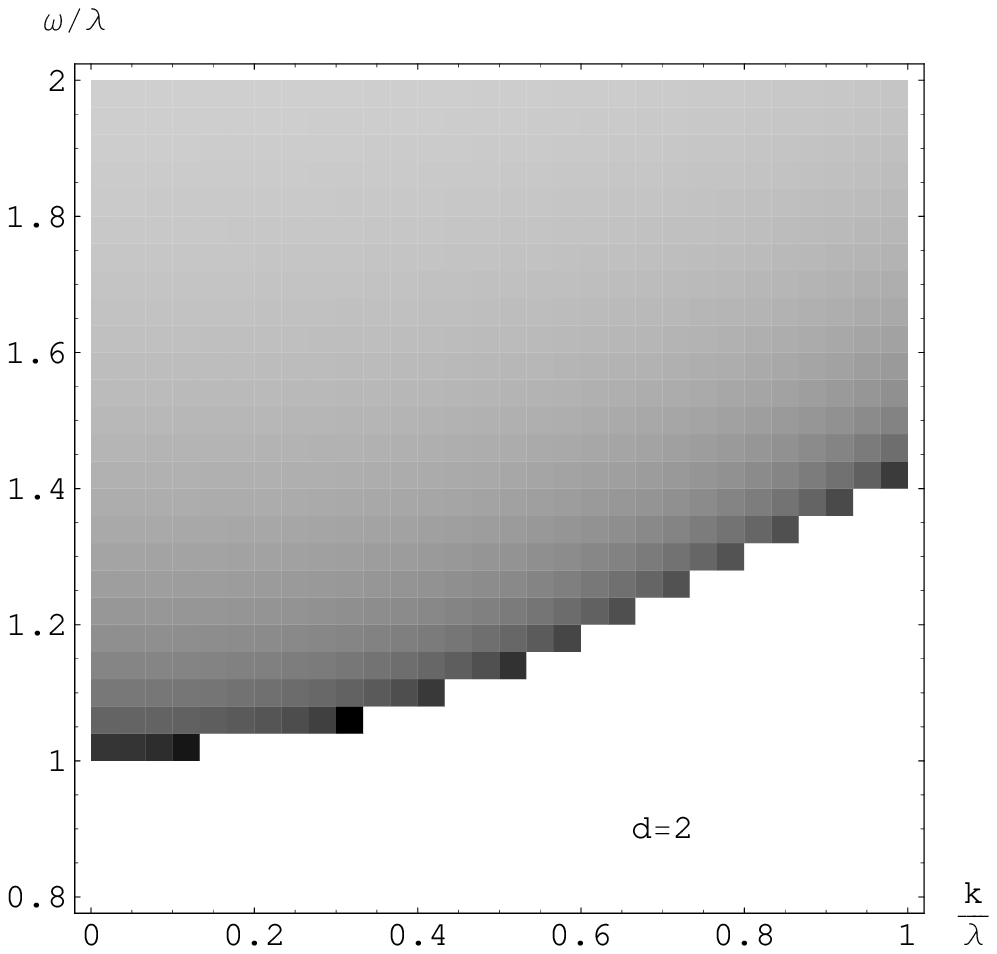}
\includegraphics{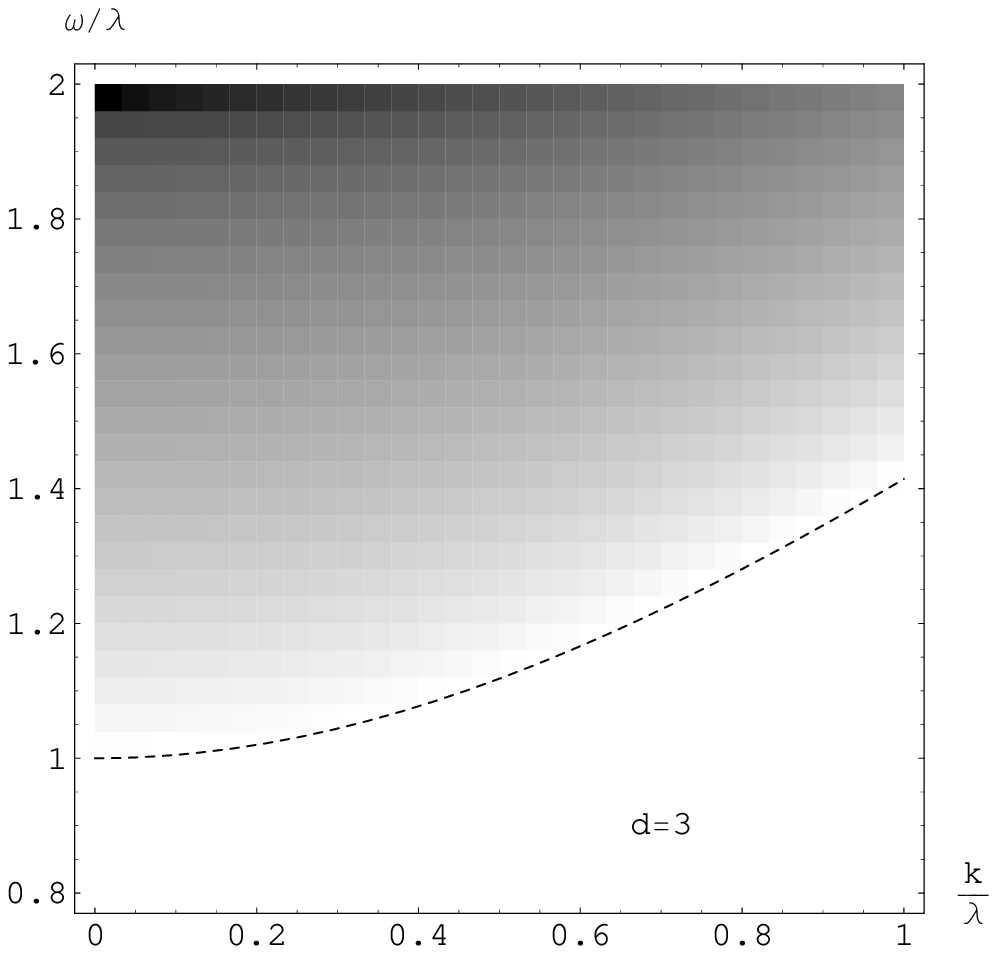}
\caption{\label{fig:SSInearSSCsigma}The imaginary part of the self
energy in perturbation theory to order $g_\psi^2$. The upper plot is
in two spatial dimensions; the lower plot is in three. In the lower
plot, the dashed line is at $\omega^2 = \lambda^2 + k^2$.}
\end{figure*}

Figure \ref{fig:SSInearSSCsigma} shows $\im \Sigma^\psi$ as a
function of $\omega$ and $k$, for $d = 2$ (upper plot) and $d = 3$
(lower plot).

\section{Critical properties: SSI--SSC (Class B2)}
\label{SectionRGd}

The transition between SSI and SSC is, as in Section
\ref{SectionRGforSSC}, described by the (imaginary-time) action \be
\Act_\Psi = \spaceitimeint \left(|\partial \Psi|^2 + r_\Psi |\Psi|^2
+ \frac{u_\Psi}{4}|\Psi|^4\right)\punc{,} \ee while the lowest-lying
spin-carrying excitations now belong to the spin-$2$ field
$d_{\mu\nu}$, with action \be \Act_{\Psi,d} = \spaceitimeint \left[
\bar{d}_{\mu\nu}\left(\partial_\tau - \frac{\nabla^2}{2m_d} + r_d
\right) d_{\mu\nu} + g_d \bar{d}_{\mu\nu}d_{\mu\nu}\bar\Psi \Psi +\,
\cdots\right]\punc{.} \ee As described in Section
\ref{SectionRegionD}, the dispersion is always irrelevant, while the
coupling $g_d$ is marginal for $d = 2$ and irrelevant above.

We therefore use perturbation theory in $g_d$ to describe the
response function $\Pi_Q$. Since $\Pi_Q$ is related to the two-point
correlator of $d_{\mu\nu}$ by \refeq{PiQfromd}, we are interested in
self-energy diagrams for $d_{\mu\nu}$. Diagrams such as \be
\Sigma_d^{(0)} = \;
\parbox{30mm}{
\begin{picture}(30,15)
\put(0,0){\includegraphics{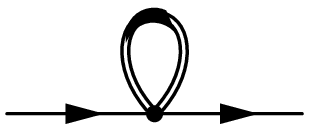}}
\fmfL(7.14854,5.39124,t){$d_{\mu \nu }$}%
\fmfL(22.85146,5.39124,t){$d_{\mu \nu }$}%

\end{picture}}\punc{,}
\ee which do not depend on the external momentum or frequency,
simply renormalize the value of $r_d$ and are of no interest. The
lowest-order diagram that we consider is therefore \be
\Sigma_d^{(1)} = \;
\parbox{30mm}{
\begin{picture}(30,20)
\put(0,0){\includegraphics{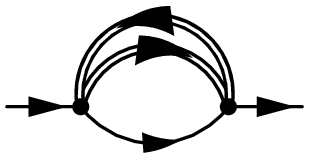}}
\fmfL(7.14854,5.39124,t){$d_{\mu \nu }$}%
\fmfL(22.85146,5.39124,t){$d_{\mu \nu }$}%

\end{picture}}\punc{.}
\ee

Since $\Psi$ is critical, rather than using the free propagator
$G^\Psi$ to describe this field, we use the full four-point
(connected) correlation function, defined by \be \Pi^\Psi(\xv,\tau)
= \Mean{\mathcal{T}_\tau \:
\bar\Psi(\xv,\tau)\Psi(\xv,\tau)\bar\Psi(\zerov,0)\Psi(\zerov,0)} -
\Mean{\bar\Psi(\zerov,0)\Psi(\zerov,0)}^2\punc{.} \ee At the
critical point, the Fourier transform of this correlation function
has a power-law form, $\Pi^\Psi(\kv,\ii\omega) = C(\sqrt{k^2 +
\omega^2})^{-y}$, where $C$ is a constant. The exponent $y$ is
related to the scaling dimensions of the operator $|\Psi|^2$ at the
critical point of $\Act_\Psi$; the latter scaling dimension is
$d+1-1/\nu$ (where $\nu$ is the correlation length exponent of the
SSI to SSC transition), and so $y=2/\nu - d - 1$. In $d=2$, the
current best value\cite{vicari} is $\nu \approx 0.6717$, and so
$y\approx -0.0225$. At leading order in $\epsilon$, a standard RG
computation shows that $y = \frac{1}{5}\epsilon +
\Order{\epsilon^2}$. (Note that using this one-loop result
corresponds to summing the diagram shown for $\Sigma_d^{(1)}$ as
well as all ladder diagrams with increasing numbers of $u_\Psi$
interactions between the two $\Psi$ propagators.)

The self-energy diagram $\Sigma_d^{(1)}$ is then given by \be
\Sigma_d^{(1)}(\ii\omega) = g_d^2 \int_{\kv,\omega}
G_0^d\biglb(\ii(\omega+\nu)\bigrb)\Pi^\Psi(\kv,-\ii\nu)\punc{,} \ee
where the free propagator for $d_{\mu\nu}$ is given by \be
G_0^d(\ii\omega) = \frac{1}{-\ii\omega + r_d}\punc{.} \ee Using the
explicit form of $\Pi^\Psi$, we have \be \Sigma_d^{(1)}(\ii\omega) =
g_d^2 C \int_\kv \int_{-\infty}^{\infty}\frac{\dd \nu}{2\pi}
\frac{1}{-\ii(\omega+\nu) + r_d}\cdot\frac{1}{(k^2 +
\nu^2)^{y/2}}\punc{.} \ee The integrand is an analytic function
apart from a pole at $\nu = - \omega - \ii r_d$ and a branch cut
along the imaginary axis for $\nu = \ii x$, where $|x| > k$.
Deforming the contour along the branch cut in the upper half-plane
gives
\begin{align}
\Sigma_d^{(1)}(\ii\omega) &= -\frac{g_d^2 C}{\pi} \int_\kv \int_{k}^{\infty}\dd x\, \frac{1}{x - \ii \omega + r_d}\im\frac{1}{(k^2 - x^2)^{y/2}}\\
&= -\frac{g_d^2 C \sin(\pi y/2)}{\pi} \int_\kv\int_{k}^{\infty}\dd
\xi\, \frac{1}{1 + \xi + (r_d- \ii
\omega)/k}\cdot\frac{1}{[k^2\xi(1+2\xi)]^{y/2}}\punc{.}
\end{align}

Now consider the quasiparticle decay rate, found by analytically
continuing $\ii\omega$ to real frequencies and taking the imaginary
part, \be \im \Sigma_d^{(1)}(r_d + \delta\omega) = -g_d^2 C \sin(\pi
y/2) \int_\kv\frac{\Theta(\delta\omega - k)}{[(\delta\omega -
k)(2\delta\omega - k)]^{y/2}}\punc{,} \ee where $\Theta$ is the
unit-step function. The frequency can now be isolated from the
integral, leaving \be \im \Sigma_d^{(1)}(r_d + \delta\omega) =
-g_d^2 C_y' \delta\omega^{d-y}\punc{,} \ee where $C_y'$ is a
function of $y$ only.

Note that $d>y$, so that the decay rate, given by $\im\Sigma_d$
evaluated at the quasiparticle energy, $\omega = r_d$, vanishes.
This is as expected, since the spin-$2$ field $d_{\mu\nu}$ is, by
assumption, the lowest-lying spin-carrying excitation. We therefore
conclude that, contrary to the case described above in Section
\ref{SectionRGforSSC}, the quasiparticle peak survives, even in the
presence of gapless critical excitations of $\Psi$.

\section{Conclusions}

This paper has described a variety of representative models of spin
dynamics across the superfluid insulator transition of spinful
bosons. The main classes can be discussed in the context of the
simple mean field phase diagram in Fig.~\ref{MeanFieldPhaseDiagram}
of $S=1$ bosons, with the three phases SSI (spin-singlet insulator),
SSC (spin-singlet condensate) and PC (polar condensate).

First, we presented the spin response spectral functions in the
three phases. The SSI and SSC phases have a spin gap, and
consequently, the associated spin spectral density $A(\omega)$ has a
non-zero value only above a threshold value given by the two
$\psi_\mu$ particle continuum; the $\psi_\mu$ particle carries
charge $Q=1$ and spin $F=1$. The nature of this threshold
singularity was presented in Eq.~(\ref{ss1}) and below, and below
Eq.~(\ref{Pi1a}) for the SSI phase. An alternative case, which could
occur only well away from the PC phase, is that a $\psi_\mu$
particle and an anti-particle form a $Q=0$ bound state with non-zero
spin, $d_{\mu\nu}$; in this case, the spin spectral function in the
SSI phase consists of a sharp quasiparticle delta function at the
spin gap energy. Closely related results apply to the SSC phase, as
discussed in Section~\ref{SSC}. Finally, for the PC phase, we have a
gapless spin excitation, leading to Goldstone spin responses noted
in Section~\ref{PC}.

Next, we described the quantum phase transitions between these
phases under the following classes:\\
({\bf A}) This is the SSI to PC transition, associated with the
condensation of the $\psi_\mu$. We found that the spin spectral
density at the quantum critical point was determined by the scaling
dimension of a composite spin operator which was bilinear in the
$\psi_\mu$, as discussed in Section~\ref{SectionRGforPC}.\\
({\bf B}) This was the transition from the SSI to the SSC phase,
driven by the condensation of $Q=2$, $F=0$ particle, $\Psi$. This
class had two subclasses.\\
({\bf B1}) The lowest non-zero spin excitation consisted of the
two-particle continuum of the $\psi_\mu$ particle. This was the most
novel case, and was discussed at length in
Section~\ref{SectionRGforSSC}. Here we found interesting fluctuation
corrections to the spin spectral density, characterized by the new
`impurity' exponent $\eta_\Pi$ in Eq.~(\ref{ss5}).\\
({\bf B2}) The spin response was associated with the sharp $Q=0$,
$F=1$ quasiparticle $d_{\mu\nu}$, which can be stable well away from
the PC phase. The damping of this quasiparticle from the critical
$\Psi$ fluctuations was found to be associated with a composite
operator whose scaling dimension was obtained in
Section~\ref{SectionRGd}. This damping led to powerlaw spectral
absorption above the spin gap, but the quasiparticle peak survived
even at the critical point.

In all of the above phases and quantum transitions we also obtained
results for the nature of the single-particle Green's function of
the $\psi_\mu$ particle. This is not directly associated with a spin
oscillation, because the $\psi_\mu$ particle has a non-zero charge
$Q$. However, this can be measured experimentally in `photoemission'
type experiments, such as microwave absorption, which involve
ejection of one boson from the atom trap.

We also did not consider the transition between the two superfluids
in Fig.~\ref{MeanFieldPhaseDiagram}, the SSC and PC. This transition
is associated with spin rotation symmetry breaking, and so has a
charge neutral, vector O(3) order parameter. Consequently, the spin
singularities can be mapped onto those of a relativistic O(3) model,
which were described in much detail in Ref.~\onlinecite{csy}.
However, here we also have to worry about the coupling of the
critical O(3) modes to the gapless `phonon' modes of the
superfluids. The nature of this coupling was discussed in a
different context in Ref.~\onlinecite{frey}, and it was found that
as long as the free energy exponent $\alpha < 0$, the coupling
between the Goldstone and critical modes was irrelevant. It is known
that this is the case for the O(3) model, and so the results of
Ref.~\onlinecite{csy} apply unchanged to the SSC to PC transition.

We thank E.~Demler, M.~Lukin, R.~Shankar, and E.~Vicari for useful
discussions. This research was supported by the NSF grants
DMR-0537077, DMR-0342157, DMR-0354517, and PHY05-51164.

\appendix

\section{Momentum cutoff RG}
\label{CutoffRG}

In Section \ref{SectionRGforSSC}, the scaling dimensions of the
particle and hole excitations across the SSI--SSC transition were
found using dimensional regularization. Here, we will peform the
same calculation to one-loop order using a momentum cutoff. (The
scaling dimension of the compound operator $p_\mu h_\nu$ can be
found by an analogous calculation.)

Our approach will be to calculate the correlation functions of the
gapped $p_\mu$ and $h_\mu$ excitations, evaluated for real
frequencies just above the gap $\lambda$. (Imaginary frequencies
will be used as a formal device when calculating the diagrams,
followed by analytic continuation.) We will find that there is a
rescaling operation that is a symmetry of the theory and relates
correlation functions evaluated at one frequency to those evaluated
at another, as in a standard RG calculation. In this case, however,
it is necessary to rescale relative to the gap energy $\lambda$,
rather than the zero of frequency.

\subsection{Self-energy renormalization}
\label{SectionSelfEnergyRG}

As a result of particle conservation, there is no one-loop diagram
contributing to the renormalization of the interaction vertex.

The only one-loop diagram for the self energy of the particle (or
hole) excitation is given in \refeq{SelfEnergyDiagram1}: \be
\Sigma^\psi_1(\ii\omega) = g_\psi^2\int_\kv
\int_{-\infty}^{\infty}\frac{\dd\omega'}{2\pi} \,
G^\Psi_0(\kv,\ii\omega')
G^\psi_0\biglb(\ii(\omega-\omega')\bigrb)\punc{,} \ee where \be
\int_\kv \equiv \int \frac{\dd^d \kv}{(2\pi)^d} = \Omega_d
\int_0^\Lambda \dd k\, k^{d-1} \ee (for an isotropic integrand),
with $\Lambda$ the cutoff. Since the dispersion of $p_\mu$ and
$h_\mu$ is irrelevant, the diagram is calculated with the external
momentum equal to zero.

Working at criticality, where we set $r_\Psi = 0$, this gives \be
\Sigma^\psi_1(\ii\omega) = g_\psi^2\int_\kv
\frac{1}{2k}\cdot\frac{1}{-\ii\omega + \lambda + k}\punc{,} \ee
after performing the integral over $\omega'$ by contour integration.
With the definition $z = -\ii\omega + \lambda$, we have \be
\Sigma^\psi_1(\ii\omega) = g_\psi^2\frac{\Omega_d}{2}\int_0^\Lambda
\dd k\, \frac{k^{d-2}}{k+z}\punc{.} \ee

Using Dyson's equation, the inverse of the propagator is therefore
\be G^\psi_1(\ii\omega)^{-1} = z - \frac{g_\psi^2
\Omega_d}{2}\int_0^\Lambda \dd k\, \frac{k^{d-2}}{k+z} +
\Order{g_\psi^4}\punc{.} \ee Taking the derivative with respect to
$\Lambda$ and expanding in powers of $z/\Lambda$ gives \be \Lambda
\parder{G_1^{-1}}{\Lambda} = -\frac{g_\psi^2
\Omega_d}{2}\Lambda^{d-2} + \frac{g_\psi^2
\Omega_d}{2}z\Lambda^{d-3} + \cdots\punc{.} \ee The first term is
independent of $\omega$ and so corresponds to a renormalization of
$\lambda$, which is of no interest to us. The second term
corresponds to wavefunction renormalization and is the only relevant
contribution from this diagram.

Since there is no diagram giving a renormalization of the coupling
$g_\psi$, a reduction in the cutoff from $\Lambda$ to
$(1-\delta)\Lambda$ (with $\delta$ infinitesimal) can be compensated
by replacing the action $\Act_{\Psi,\psi}$ by
\begin{multline}
\label{ChangeInAction}
\Act_{\Psi,\psi} + \delta\Act_{\Psi,\psi} = \spaceitimeint \bigg\{\left(1+\hat{g}_\psi^2\delta\right) \left[ \bar{p}_\mu \left(\partial_\tau + \lambda\right) p_\mu + \bar{h}_\mu \left(\partial_\tau + \lambda\right) h_\mu \right]\\
+ g_\psi \left( \bar\Psi p_\mu \bar{h}_\mu + \Psi \bar{p}_\mu h_\mu
\right)\bigg\}\punc{.}
\end{multline}
To simplify this expression slightly, we have defined the
dimensionless quantity\footnote{Note the similarity to the
corresponding definition in \refeq{Definegtilde}, since
$\Omega_3^{-1} = 2\pi^2$.} $\hat{g}_\psi = g_\psi \Lambda^{-(d-3)/2}
\sqrt{\Omega_d/2}$.

\subsection{Partition function}

\newcommand{\Z}{\mathcal{Z}}
\newcommand{\doubleint}{\int\!\!\!\!\int}

This notion of `compensating a reduction in the cutoff' can be made
more precise by considering the partition function with discrete
sources:
\begin{multline}
\label{DefineZ}
\Z_\Lambda(J_i,\kv_i,z_i,g_\psi) = \int_\Lambda \D^2p\,\D^2h\:\exp -\bigg\{ \int_{\omega,\kv}(\bar{p}_\mu z p_\mu + \bar{h}_\mu z h_\mu)\\
+ g_\psi\int_{\omega_1,\kv_1} \int_{\omega_2,\kv_2} \left[\bar\Psi\biglb(\kv_1-\kv_2,\ii(\omega_1-\omega_2)\bigrb) p_\mu(\kv_1,\ii\omega_1) \bar{h}_\mu(\kv_2,\ii\omega_2) \pcc\right] \\
+ \sum_i \bar{J}_{i\mu} [p_\mu(\kv_i,\lambda -
z_i)+\bar{h}_\mu(\kv_i,\lambda - z_i)]\pcc\bigg\}\punc{,}
\end{multline}
from which correlation functions can be found by successive
differentiation with respect to $J_i$ and $\bar{J}_i$. (We are
concerned with $\psi_\mu$, so integration over $\Psi$, with the
appropriate measure, is implied.) The subscript $\Lambda$ on the
integral sign denotes that a cutoff $\Lambda$ should be used.

Using this definition, \refeq{ChangeInAction} can be written
\begin{multline}
\Z_\Lambda(J_i,z_i,g_\psi) = \int_{(1-\delta)\Lambda} \D^2p\,\D^2h\:\exp -\bigg\{ \left(1+\hat{g}_\psi^2\delta\right)\int_{\omega,\kv}(\bar{p}_\mu z p_\mu + \bar{h}_\mu z h_\mu)\\
+ g_\psi\doubleint_{\omega,\kv} \left(\bar\Psi p_\mu \bar{h}_\mu
\pcc\right) + \sum_i \bar{J}_{i\mu} [p_\mu(\lambda -
z_i)+\bar{h}_\mu(\lambda - z_i)] \pcc\bigg\}\punc{,}
\end{multline}
which expresses the fact that the partition function, and hence all
correlators, are unchanged by a shift in the cutoff and a
compensating change in the action. A condensed notation has been
used, where the momentum dependence is suppressed throughout.

To bring this closer to the form of \refeq{DefineZ}, we rescale the
fields $p_\mu$ and $h_\mu$ within the functional integral. By doing
so, we can return the coefficient of the quadratic term to unity,
giving
\begin{multline}
\Z_\Lambda(J_i,z_i,g_\psi) = \int_{(1-\delta)\Lambda} \D^2p\,\D^2h\:\exp -\bigg\{ \int_{\omega,\kv}(\bar{p}_\mu z p_\mu + \bar{h}_\mu z h_\mu)\\
+ g_\psi\left(1-\hat{g}_\psi^2\delta\right)\doubleint_{\omega,\kv} \left(\bar\Psi p_\mu \bar{h}_\mu \pcc\right)\\
+ \sum_i \left(1-\frac{\hat{g}_\psi^2}{2}\delta\right)J_{i\mu}
[p_\mu(\lambda - z_i)+\bar{h}_\mu(\lambda - z_i)]\pcc\bigg\}\punc{.}
\end{multline}
By comparison with \refeq{DefineZ}, we can write \be \label{RGStep1}
\Z_\Lambda(J_i,\kv_i,z_i,g_\psi) =
\Z_{(1-\delta)\Lambda}\biglb(\left(1-\hat{g}_\psi^2\delta/2\right)J_i,
\kv_i, z_i, \left(1-\hat{g}_\psi^2\delta\right)g_\psi\bigrb)\punc{.}
\ee

\subsection{Rescaling}

To return to the original theory, with cutoff $\Lambda$, we now
perform a rescaling of all variables according to their engineering
dimensions, with $[\tau] = [\xv] =-1$. Since we are working at the
critical point of $\Psi$, we have \be \Psi(\xv/b, \tau/b) =
b^X\Psi(\xv, \tau)\punc{,} \ee where $X$ is the scaling dimension of
the field $\Psi$. By dimensional analysis of
\refeq{PsiCriticalTheory}, the engineering dimension of $\Psi$ is
seen to be $[\Psi] = (D-2)/2 = (d-1)/2$ and one would naively expect
$X = [\Psi] = 1 - \frac{\epsilon}{2}$. This expectation actually
happens to be correct (to order $\epsilon$), since there is no
wavefunction renormalization of $\Act_\Psi$ to one-loop order.

Performing this rescaling leads to \be \label{RGStep2}
\Z_\Lambda(J_i,\kv_i,z_i,g_\psi) =
\Z_{b\Lambda}(b^{1+d/2}J_i,b\kv_i,bz_i,b^{1-X}g_\psi)\punc{,} \ee
after making the substitutions $\psi'(\kv, \lambda-z) = b^{d/2 + 1}
\psi(b\kv, \lambda - bz)$ and $\lambda - \ii\omega' = (\lambda -
\ii\omega) / b$.

This can now be combined with \refeq{RGStep1} to give
\begin{multline}
\label{RGResult} \Z_\Lambda(J_i,\kv_i,z_i,g_\psi) =
\Z_\Lambda\biglb((1-\hat{g}_\psi^2\delta/2)(1+(1+d/2)\delta)J_i, (1+\delta)\kv_i, (1+\delta)z_i,\\
(1-\hat{g}_\psi^2\delta)(1+(1-X)\delta)g_\psi\bigrb)\punc{.}
\end{multline}
This gives a relationship between correlators in the same theory but
at different frequencies.

The fixed point of \refeq{RGResult} occurs when \be g_\psi^\star =
[1-(\hat{g}_\psi^\star)^2\delta][1+(1-X)\delta]g_\psi^\star\punc{,}
\ee so that \be \hat{g}_\psi^\star = \sqrt{1-X} =
\sqrt{\frac{\epsilon}{2}}\punc{,} \ee which should be compared with
\refeq{FixedPoint}.

\subsection{Renormalized propagator}

At the fixed point, \refeq{RGResult} becomes \be
\Z_\Lambda(J_i,\kv_i,z_i,g_\psi^\star) =
\Z_\Lambda\biglb((1+y\delta)J_i, (1+\delta)\kv_i, (1+\delta)z_i,
g_\psi^\star\bigrb)\punc{,} \ee where $y =
1+d/2-(\hat{g}_\psi^\star)^2/2 = 5/2 - 3\epsilon/4$. Taking
derivatives with respect to $J_\mu$ and $\bar{J}_\nu$ gives
\begin{multline}
\label{ScalingOfCorrelator}
\Mean{\bar{\psi}_\mu(\kv_1,\lambda - z_1)\psi_\nu(\kv_2,\lambda - z_2)} =\\
(1+2y\delta)\big\langle\bar{\psi}_\mu\biglb((1+\delta)\kv_1,\lambda
- (1+\delta)z_1\bigrb)\psi_\nu\biglb((1+\delta)\kv_2,\lambda -
(1+\delta)z_2\bigrb)\big\rangle\punc{.}
\end{multline}

Using the conservation of frequency and momentum, we can define the
propagator $G^\psi$ by \be \Mean{\bar{\psi}_\mu(\kv_1,\lambda -
z_1)\psi_\nu(\kv_2,\lambda - z_2)} = (2\pi)^{d+1}\delta^d(\kv_1 -
\kv_2) \delta(z_1 - z_2) \delta_{\mu\nu} G^\psi(\kv_1,\lambda -
z_1)\punc{.} \ee Using this definition, \refeq{ScalingOfCorrelator}
becomes \be G^\psi(\kv, \lambda - z) =
(1+y'\delta)G^\psi\biglb((1+\delta)\kv, \lambda -
(1+\delta)z\bigrb)\punc{,} \ee with $y' = 2y - d - 1 = 1 -
\epsilon/2$.

Restricting attention to $\kv = \zerov$ gives \be G^\psi(\zerov,
\lambda - z) = (1+y'\delta)G^\psi\biglb(\zerov, \lambda -
(1+\delta)z\bigrb)\punc{,} \ee which can be iterated to give \be
G^\psi(\zerov, \lambda - z) \sim z^{-1 + \epsilon/2}\punc{.} \ee
Equivalently, after analytic continuation to real frequencies, we
have \be G^\psi(\zerov, \omega) \sim (\lambda -
\omega)^{-1+\epsilon/2}\punc{,} \ee which agrees with
\refeq{TwoPointCorrelatorResult}.

\end{document}